\documentclass[aps,prc,twoside,twocolumn,nofootinbib,%
showpacs,showkeys,floatfix,dvips]{revtex4}

\usepackage{amsmath,amssymb,bm}
\usepackage{graphicx}

\usepackage[normalem]{ulem} 
\usepackage[dvips]{color} 
\renewcommand\sout{\bgroup \color{red} \ULdepth=-.5ex \ULset}

\newcommand*{\fs}[1]{{#1\!\!\!/}}
\newcommand*{\hc}{\text{ H.\,c.}}

\begin{document}

\title{\boldmath Combined Analysis of $\eta$ Meson
Hadro- and Photo-production off Nucleons}

\author{K. \surname{Nakayama}}
\email{nakayama@uga.edu}

\affiliation{Department of Physics and Astronomy, University of Georgia,
Athens, GA 30602, USA}
\affiliation{Institut f{\"u}r Kernphysik (Theorie),
Forschungszentrum J{\"u}lich, 52425 J{\"u}lich, Germany}

\author{Yongseok \surname{Oh}}
\email{yohphy@knu.ac.kr}

\affiliation{Department of Physics, 
Kyungpook National University, Daegu 702-701, Korea}

\author{H. \surname{Haberzettl}}
\email{helmut.haberzettl@gwu.edu}

\affiliation{Center for Nuclear Studies, Department of Physics,
The George Washington University, Washington, DC 20052, USA}

\date{\today}

\begin{abstract}
The $\eta$-meson production in photon- and hadron-induced reactions, namely,
$\gamma p \to p \eta$, $\pi^- p \to n \eta$, $pp \to pp\eta$, and $pn \to
pn\eta$, are investigated in a combined analysis in order to learn about the
relevant production mechanisms and the possible role of nucleon resonances in
these reactions.
We consider the nucleonic, mesonic, and nucleon resonance currents
constructed within an effective Lagrangian approach and compare the
results with the available data for cross sections and spin asymmetries for
these reactions.
We found that the reaction $\gamma p \to p \eta$ could be described well with
the inclusion of the well-established $S_{11}(1535)$, $S_{11}(1650)$,
$D_{13}(1520)$, and $D_{13}(1700)$ resonances, in addition to the mesonic
current.
Consideration of other well-established resonances in the same mass region,
including the spin-5/2 resonances, $D_{15}(1675)$ and $F_{15}(1680)$, does
not further improve the results qualitatively.
For the reaction $\pi^- p \to n \eta$, the $P_{13}(1720)$ resonance is found
to be important for reproducing the structure observed in the differential
cross section data.
Our model also improves the description of the reaction $NN \to NN\eta$ to
a large extent compared to the earlier results by 
Nakayama \textit{et al.\/} [Phys.\ Rev.\ C \textbf{68}, 045201 (2003)].
For this reaction, we address two cases where either the $S_{11}(1535)$ or 
the $D_{13}$ dominates.
Further improvement in the description of these reactions and the difficulty
to uniquely determine the nucleon resonance parameters in the present type of
analysis are discussed.
\end{abstract}

      

\maketitle


\section{Introduction}
\label{sec:introduction}

One of the primary motivations for studying the production of mesons off
nucleons is to investigate the structure and the properties of nucleon resonances
and, in the case of heavy-meson productions, to learn about hadron dynamics at
short range.
In particular, a clear understanding of the production mechanisms of mesons
heavier than the pion still requires further theoretical and experimental
investigation.
Apart from pion production, the majority of theoretical investigations of
meson-production processes are performed within phenomenological
meson-exchange approaches.
Such an approach forces one to correlate as many independent processes as
possible within a single model if one wishes to extract meaningful physics
information.
Indeed, this is the basic motivation behind the coupled-channels approaches.

In this paper, we present the result of our investigation of $\eta$-meson
production in both  photon- and hadron-induced reactions, which include
\begin{align}
\gamma + N & \to  N + \eta  ,\nonumber  \\
\pi + N & \to  N + \eta  , \label{eq:1}\\
N + N & \to  N + N + \eta  .\nonumber
\end{align}
More specifically, we perform a combined analysis of the reactions, $\gamma p
\to p \eta$, $\pi^- p \to n \eta$, $pp \to pp\eta$, and $pn \to pn\eta$ based
on an effective Lagrangian approach.

The amount of data available for $\eta$-meson production is now considerable.
In particular, in photoproduction processes off protons, a new generation of
high-precision data is now available, not only for total and differential
cross sections in a wide range of energies starting from
threshold~\cite{photo-xsc-data,photo-xsc-data1}
but also for beam and target
asymmetries~\cite{photo-spin-data}.
Taken in combination, these data sets offer
better opportunities for investigating the properties of nucleon resonances. In
particular, much more detailed studies than in the past are possible now for
resonances that may perhaps couple strongly to $N\eta$, but only weakly to
$N\pi$. In this respect, the recent data on the quasi-free $\gamma n \to n
\eta$ process~\cite{KCGG07} have attracted much interest in $\eta$-production
processes in connection to the possible existence of a narrow (crypto-exotic)
baryon resonance with a mass near $1.68$~GeV, which is still under
debate~\cite{exotic}.
(See also Refs.~\citenum{LNS,CB-ELSA,anti-dec}.)

There are a large number of theoretical investigations of $\eta$
photoproduction, mostly from the early 1990s to the
present~\cite{photo-theory}
and, especially, off protons. Most of them focus on the role of nucleon
resonances and the extraction of the corresponding resonance parameters, but a
variety of issues have also been addressed, such as the $NN\eta$ coupling
constant~\cite{TBK94}, the $U_A(1)$ anomaly~\cite{BWW00}, and the extended
chiral symmetry~\cite{FMU06}. Among the more recent calculations that
analyze the recent high-precision data are those of
Refs.~\citenum{MAID,ASBK05,CQM}.
The $\eta$-MAID approach~\cite{MAID} is an isobar model that includes, in
addition to the $t$-channel vector-meson exchange or its Reggeized version, a
set of well-established spin-1/2, -3/2, and -5/2 resonances. This model has
been applied to the analyses of data for photoproduction as well as for
electroproduction of the $\eta$ meson.
The Bonn-Gatchina-Giessen group~\cite{ASBK05} has
developed a model which has been employed in a combined partial wave analysis
of the photoproduction data with $N\pi$, $N\eta$, $\Lambda K$, and $\Sigma K$
final states. This model considers fourteen nucleon resonances and seven Delta
resonances for achieving a reasonable overall agreement with the whole database
considered in their analysis.%
\footnote{The $\Delta$ resonances do not couple to the $N\eta$ and the $\Lambda K$
channels because of isospin conservation.}
He, Saghai, and Li have analyzed the $\eta$-photoproduction reaction in a chiral
constituent quark model~\cite{CQM} by considering all of the one- to 
four-star-rated resonances listed in the Review of Particle Data Group
(PDG)~\cite{PDG06}.

In hadronic reactions, a noticeable amount of data for $NN \to NN\eta$ has
been
accumulated~\cite{NNeta-xsc-data,COSY-TOF-02,COSY11,CELSIUS-WASA,NNeta-spin-data}.
Here, we have $pp$ and $p\eta$ invariant mass distributions and the analyzing
power near threshold, in addition to differential and total cross sections.
The $NN \to NN\eta$ process is particularly relevant for studying the role
of the $N\eta$ final-state interaction (FSI).
Most of the existing calculations (see, e.g.,
Refs.~\citenum{NSL02}, \citenum{Hanhart03} and
references therein; see also Ref.~\citenum{CSZ06})
take into account the effects of the $NN$ FSI in one way or another,
which is well-known to influence the energy dependence of the cross section
near threshold.
Calculations that include the $N\eta$ FSI to lowest
order~\cite{BGHH03}
reproduce the bulk of the energy dependence exhibited by the data.
However, they are not sufficient to reproduce the $pp$ invariant mass
distribution measured by the COSY-TOF~\cite{COSY-TOF-02} and COSY-11
Collaborations~\cite{COSY11}.
Thus, in order to explain the observed $pp$ invariant mass distribution, the
importance of the three-body nature of the final state (in the $S$-wave) has
been emphasized~\cite{FA03} or an extra energy dependence in the basic
production amplitude has been suggested~\cite{Deloff03}.
In spite of this, another possibility has been offered, which is based on the higher partial
wave ($P$-wave) contribution~\cite{NHHS03}.
We observe that what is actually required to reproduce the measured $pp$
invariant mass distribution is an extra $p^{\prime 2}$ dependence, where
$p^\prime$ denotes the relative momentum of the final $pp$ subsystem.
Obviously, this can be achieved either by an $S$-wave or by a $P$-wave
contribution.
Note that the $NN$ $P$-wave (${}^3P_0$) can also yield a flat proton angular
distribution, as observed in the corresponding data.
The model of Ref.~\citenum{NHHS03}, however, underpredicts the measured total cross
near threshold to a large extent.
One of the objectives of the present study is to resolve this discrepancy.
In any case, as pointed out in Ref.~\citenum{NHHS03}, the measurements of the
spin-correlation functions should help settle the question of the $S$-wave
versus the $P$-wave contributions in a model-independent way.

Most data for the more basic (two-body) $\pi N \to N \eta$ reaction have been
obtained in the 1960s through 1970s; they are rather scarce and less
accurate~\cite{piNetaN-data}
than the data for the other two reactions mentioned above.
Recently, the Crystal Ball Collaboration has measured the differential and the
total cross sections of this reaction near threshold~\cite{CB05}.
Theoretically, this reaction has been studied mostly in conjunction with other
reactions in a combined analysis~\cite{BT90} or in coupled-channels
approaches~\cite{IOV01,PM02c,GHHS03,CSZ06a,MSL06} in order to constrain
some of the model parameters. Recently, Zhong \textit{et
al.\/}~\cite{ZZHS07}
extended the chiral constituent quark model for meson
photoproduction~\cite{CQM} to this reaction. Also, $\pi N \to N \eta$ has been
studied within a heavy-baryon chiral perturbation theory
\cite{Krippa01}. Arndt
\textit{et al.\/}~\cite{ABMS05}  investigated the role of $\pi N \to N \eta$
on the $N\eta$ scattering length within a coupled-channels analysis of this
reaction and  elastic $\pi N$ scattering.

In the present work, we consider the three reactions mentioned above in the
following manner: The photoproduction reaction is calculated by considering the
$s$-, $u$- and $t$-channel Feynman diagrams plus a generalized contact
term~\cite{NH04-NH05}, which ensures gauge invariance of the total amplitude, in
addition to accounting for the final-state interaction effects. (See
Ref.~\citenum{HNK06} for details.)
The $\pi N \to N \eta$ reaction is calculated in the tree-level approximation,
including the $s$-, $u$-, and $t$-channels.
To the extent that this reaction is dominated by the excitation of the
$S_{11}(1535)$ resonance at least for energies close to the threshold, this
should be a reasonable approximation if we confine ourselves to energies not
too far from the threshold.
The $N N\to N N \eta$ process is calculated in the distorted-wave Born
approximation (DWBA), where both the $NN$ FSI and the initial-state
interaction (ISI) are taken into account explicitly~\cite{NSL02}.
The $NN$ FSI is known to be responsible for the dominant energy dependence
observed in the total cross section (apart from the dependence due to the phase
space) arising from the very strong interaction in the $S$-wave states at very
low energies~\cite{NSHHS98}.
As for the basic meson-production amplitude, our model includes the nucleonic,
mesonic, and nucleon resonance currents, which are derived from the relevant
effective Lagrangians.

Ultimately, our goal is to perform a more complete model calculation in which the
relevant FSIs are taken into account explicitly. However, before being able to
undertake a complex calculation that couples many channels, we need to learn
some of the basic features of meson production (in particular, those of
$\eta$-meson production) within a simplified model where these basic features
may be revealed and analyzed in a much easier manner. In this regard, one of
the major purposes of the present investigation is to show that consideration
of the hadronic reactions $NN \to NNM$, in conjunction with more basic two-body
reactions, would greatly help in the study of nucleon resonances, especially,
in imposing much stricter constraints on the extracted resonance-nucleon-meson
($RNM$) coupling strength involving a meson $M$ other than the pion. In fact,
currently, our knowledge of the branching ratios of the majority of the
known resonances is very limited~\cite{PDG06}.

This paper is organized as follows.
In the next section, we briefly describe our model for the reactions listed
in Eq.~(\ref{eq:1}).
The results of the corresponding model calculations are presented and
discussed in Sec.~\ref{sec:results}.
Section~\ref{sec:summary} contains a summary and a conclusion.
Some details of the present model are given in the Appendix.


\section{Reaction Mechanisms}
\label{sec:model}

\begin{figure}[t!]\centering
\includegraphics[width=0.9\columnwidth,angle=0,clip]{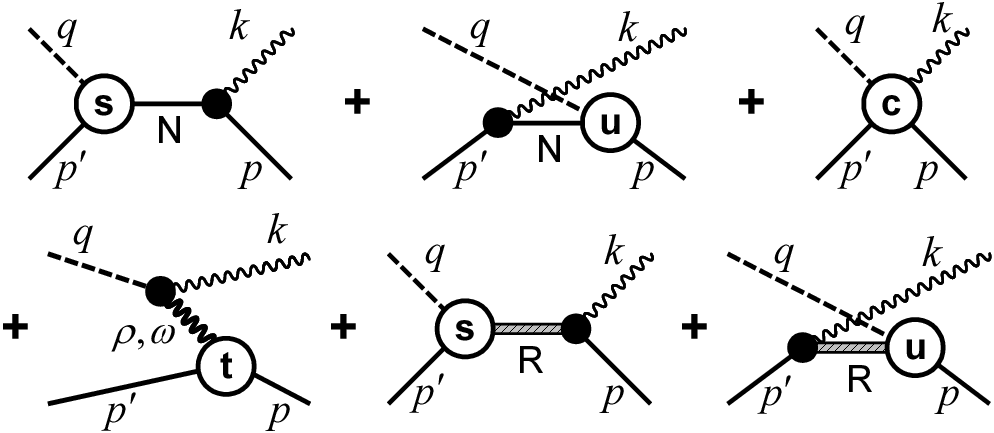}
\caption{\label{fig:diagram_photo}%
Feynman diagrams contributing to $\gamma p \to p \eta $.
Time proceeds from right to left.
The wavy, solid, and dashed lines represent the photon, the nucleon, and
the $\eta$ meson, respectively.
The intermediate baryon states are denoted by \textsf{N} and \textsf{R} for
the nucleon and the nucleon resonances, respectively.
The intermediate mesons in the $t$-channel include the $\rho$ and the $\omega$.
The external legs are labeled by the four-momenta of the respective particles,
and the labels \textsf{s}, \textsf{u}, and \textsf{t} of the hadronic vertices
correspond to the off-shell Mandelstam variables of the respective
intermediate particles.
The top-right diagram is the generalized contact current.
}
\end{figure}

\begin{figure}[t!]\centering
\includegraphics[width=0.9\columnwidth,angle=0,clip=]{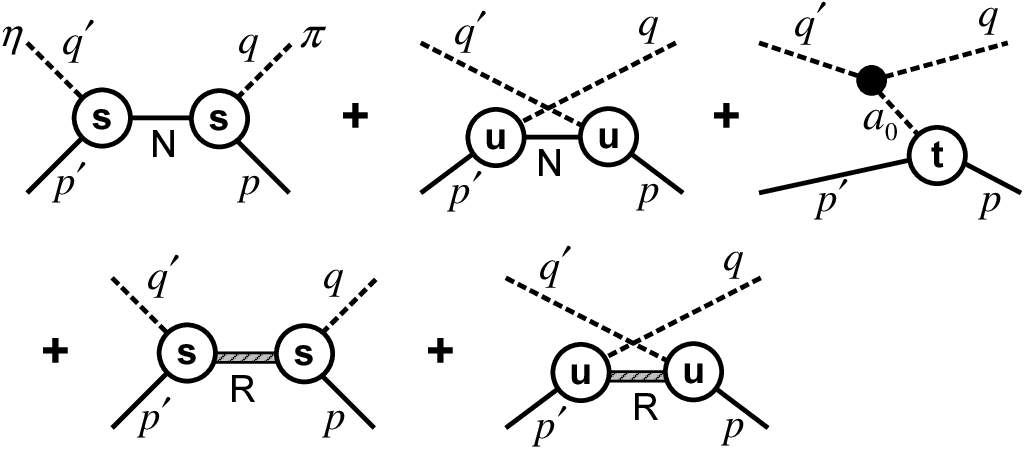}
\caption{\label{fig:diagram_pieta}%
Feynman diagrams contributing to $\pi^- p \to n \eta $.
The notation is the same as in Fig.~\ref{fig:diagram_photo}.
}
\end{figure}

\begin{figure}[t!]\centering
\includegraphics[width=0.9\columnwidth,angle=0,clip]{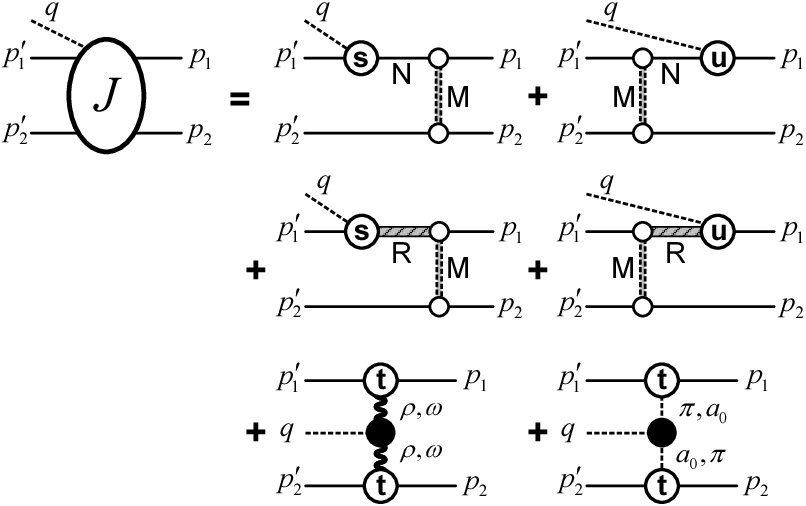}
\caption{\label{fig:diagram_NNeta}%
Basic production amplitude for $NN\to NN \eta$.
The full amplitude, including the $NN$ ISI and FSI contributions, is given
by Eq.~(\ref{eq:NNeta_ampl}).
As in Fig.~\ref{fig:diagram_photo},
\textsf{N} and \textsf{R} denote the intermediate nucleon and resonances,
respectively, and \textsf{M} incorporates all exchanges of mesons $\pi$,
$\eta$, $\rho$, $\omega$, $\sigma$, and $a_0^{}$ (former $\delta$) for the
nucleon graphs and $\pi$, $\eta$, $\rho$, and $\omega$ for the resonance
graphs.
External legs are labeled by the four-momenta of the respective particles
as in Fig.~\ref{fig:diagram_photo}.
Diagrams with $p_1^{} \leftrightarrow p_2^{}$ are understood although
not displayed here.
 }
\end{figure}

In the present work, the $\eta$-meson production processes are treated within a
relativistic meson-exchange approach, whose dynamical contents are summarized
by the Feynman diagrams displayed in Figs.~\ref{fig:diagram_photo},
\ref{fig:diagram_pieta}, and \ref{fig:diagram_NNeta} for the reactions
$\gamma N \to N \eta$, $\pi N \to N \eta$, and $N N \to N N \eta$,
respectively.
We employ phenomenological form factors at hadronic vertices to account for the
structures of the corresponding hadrons.

For the photoproduction process, the total amplitude in the present work is
given by the Feynman diagrams displayed in Fig.~\ref{fig:diagram_photo}. 
In these Feynman diagrams, the three diagrams in the lower part are transverse
individually while the three diagrams in the upper part are not.
Gauge invariance of the total amplitude is ensured by the generalized contact current
given in Refs.~\citenum{NH04-NH05} and \citenum{HNK06}, 
which follows the general formalism of
Refs.~\citenum{Habe97,HBMF98a,DW01a}. This contact
term provides a rough phenomenological description of the FSI and is not
treated explicitly here~\cite{HNK06}. The details of the present approach are
fully described in Ref.~\citenum{NH04-NH05}, where $\eta^\prime$-meson production in
photon- and hadron-induced reactions was investigated, and they will not be
repeated here. One new feature in the present work, however, is the inclusion
of spin-5/2 resonance contributions.
For our discussion, we refer the nucleonic current (NUC) to the diagrams shown in
the top line of Fig.~\ref{fig:diagram_photo}. The meson-exchange current (MEC) 
and the resonance current contributions correspond to the leftmost diagram
and the two diagrams on the right of the bottom line of Fig.~\ref{fig:diagram_photo},
respectively.

As mentioned in Sec.~\ref{sec:introduction}, the reaction $\pi N \to N \eta$
is calculated within the tree-level approximation.
To the extent that it is dominated by the $S_{11}(1535)$ resonance
contribution, this is a reasonable approximation, at least near threshold.
The total amplitude for this reaction is, therefore, given by the Feynman
diagrams displayed in Fig.~\ref{fig:diagram_pieta}.
Here, the nucleonic current (NUC) corresponds to the
first two diagrams on the top line while the meson-exchange current (MEC)
and the resonance current contributions correspond, respectively, to the
rightmost diagram on the top line and the two diagrams on the
bottom line in Fig.~\ref{fig:diagram_pieta}.

As for the $N N \to N N \eta$ reaction, the total amplitude is calculated
within the DWBA:
\begin{equation}
M = (T_f G_f + 1) J (1 + G_i T_i)  ,
\label{eq:NNeta_ampl}
\end{equation}
where $T_{i(f)}$ denote the $NN$ ISI (FSI),
$G_{i(f)}$ stands for the corresponding $NN$ propagator, and
$J$ denotes the basic production amplitude displayed in
Fig.~\ref{fig:diagram_NNeta} and is constructed from the interaction
Lagrangians given in Appendix.
Further details of the present approach to this reaction, including
all the values of the coupling constants and the cutoff parameters of the 
corresponding form factors that enter in the definition of the basic 
production amplitude $J$, can be found in Ref.~\citenum{NSL02}.
Also, we use the $NN$ interaction based on the Paris
potential~\cite{LLRV80},
which includes the Coulomb interaction as well~\cite{NADS00}.
The nucleonic, resonance, and meson-exchange contributions
correspond, respectively, to the first, second, and third lines of the
Feynman diagrams on the right-hand side in Fig.~\ref{fig:diagram_NNeta}.

In the Appendix, we present all the hadronic and electromagnetic
interaction Lagrangians and propagators necessary for computing the
diagrams displayed in Figs.~\ref{fig:diagram_photo},
\ref{fig:diagram_pieta}, and \ref{fig:diagram_NNeta} within the
present approach.
The phenomenological form factors used in this model are also given in the
Appendix.
The free parameters of our model --- the resonance parameters,
the $NN\eta$ coupling constant, and the cutoff parameter $\Lambda^*_v$ at
the electromagnetic vector-meson exchange vertex --- are fixed
so as to reproduce the available data in a global fitting procedure of
the three reaction processes listed in Eq.~(\ref{eq:1}).

\section{Results and Discussion}\label{sec:results}

In this section, we present and discuss the results of our model calculation.
The basic strategy of our approach is, in principle, the same as that of
Ref.~\citenum{NH04-NH05}; namely, we start with the nucleon plus meson-exchange 
currents and add resonance contributions one by one as needed in the fitting 
procedure until achieving a reasonable description of the available experimental 
data for the reactions listed in Eq.~(\ref{eq:1}). 
Apart from the dominant $S_{11}(1535)$ resonance, we allow for other 
well-established resonances of spin-$1/2$, -$3/2$, and -$5/2$ in this model. 
We confirm the earlier finding~\cite{photo-theory} that, in photoproduction, 
in addition to spin-$1/2$ resonances, at least spin-3/2 resonances 
--- in particular, $D_{13}$ resonances --- are needed in order to obtain a reasonable 
description of the data.

Following Ref.~\citenum{NH04-NH05}, for each resonance, we take into account only the
branching ratios $\beta_{N\pi}^{}$, $\beta_{N\eta}^{}$, and
$\beta_{N\pi\pi}^{}$ corresponding to the respective hadronic decay channels $R
\to N\pi$, $R \to N\eta$, and $R \to N \pi \pi$. The latter accounts
\emph{effectively\/} for all the other open decay channels. Note that the
branching ratios $\beta_{N\pi}^{}$ and $\beta_{N\eta}^{}$ are related to the
corresponding $RN\pi$ and $RN\eta$ coupling constants in the interaction
Lagrangians, and, as such, they are \emph{not\/} free parameters of the model.
The same is true for the branching ratio $\gamma_{N\gamma}^{}$, associated with
the radiative decay channel, which is related to the corresponding $RN\gamma$
coupling constants. Then, in view of the constraint given by
Eq.~(\ref{eq:brsum}), the branching ratio $\beta_{N\pi\pi}^{}$ is not a free
parameter either. Here, we emphasize that, in contrast to
Ref.~\citenum{NH04-NH05} ---
where some assumptions were made concerning the values of the branching ratios
$\beta_{N\pi}^{}$ --- no such assumptions are enforced in the present work
because the simultaneous consideration of the reaction processes listed in
Eq.~(\ref{eq:1}) allows us, in principle, to extract the $RN\pi$ and
$RN\eta$ coupling constants separately.

The coupling constants of the $RNV$ ($V=\rho ,\omega$) interaction
Lagrangians are required in the calculation of the $NN\to NN\eta$ reaction.
Therefore, in principle, the hadronic branching ratios involving vector mesons 
(such as the $\rho$ and the $\omega$) should also be taken into account. 
However, since we have restricted ourselves to nucleon resonances for which the
decay channels $R\to NV$ are either closed or nearly closed, we have set the
associated branching ratios to zero ($\beta_{NV}^{}=0$).%
\footnote{Strictly speaking, even these resonances can have non-vanishing
branching ratios to $NV$ channels due to their large widths.}
Obviously, in a more refined calculation, this condition should be relaxed.
In a more complete (coupled-channels) dynamical model approach, the branching
ratios and total widths will be generated by the model via the dressing
mechanism of the corresponding vertices and resonance masses.

The results shown here are not necessarily the best fits 
achievable within the present approach.
Rather, they are sample fits that illustrate different dynamical
features that may be obtained in this type of analysis.
The resonance parameters are obtained by global fitting to the available data
for the reactions mentioned above.

For this end, we consider four models.
Although these four models contain the same nucleonic and mesonic currents 
described in Figs.~\ref{fig:diagram_photo}, \ref{fig:diagram_pieta}, and
\ref{fig:diagram_NNeta}, they include different nucleon resonances and different 
resonance parameters.
In model (A), we consider only the $S_{11}(1535)$, $S_{11}(1650)$,
$D_{13}(1520)$, and $D_{13}(1700)$ resonance currents.
Model (B) includes the same resonances as in model (A), but the parameters of those
resonances are different from those of model (A). The parameters are obtained by
using different starting values for the search in parameter space during the global
fit procedure. The implication of the differences between these two models will be 
discussed later.
In addition to the resonances considered in model (A) and (B), model (C) includes
the $P_{13}(1720)$ resonance.
Finally, we consider model (D), which takes into account the contributions from
$D_{15}(1675)$, $F_{15}(1680)$, and $P_{11}(1710)$ in addition to the resonances
considered in model (C).
In the following, we present and discuss the results for each reaction.

%
%
\begin{table*}[t]\centering
\caption{\label{tbl:photo4c}
Parameters of model (A) fitted to the reactions listed in Eq.~(\ref{eq:1}).
(See the Appendix for an explanation of the parameters.) Values in boldface are
\emph{not\/} fitted. The branching ratios $\gamma_{N\gamma}^{}$ and
$\beta_{Nj}^{}$ ($j=\pi, \eta, \pi\pi$) are not free parameters, but are
extracted from the corresponding coupling constants, except for
$\beta_{N\pi\pi}^{}$ which is obtained from Eq.~(\ref{eq:brsum}). The values in
square brackets are the range estimates quoted in PDG~\cite{PDG06}. The data set
for $\gamma p \to p \eta$, $\pi^- p \to n \eta$, and $NN \to NN\eta$ was used
in the fit.}
\begingroup
\begin{tabular}{l@{\qquad}r@{\qquad}l@{\qquad}r@{\qquad}l}
\hline\hline
Nucleonic current: & & & & \\
($g_{NN\eta}^{}$, $\lambda$)                & ($0.007$,
$\textbf{0.0}$)           &      &      &        \\
\hline
Mesonic current: & & & &  \\
$\Lambda^*_v$ (MeV)   &  $1168$     & &  &  \\
\hline
$N_{11}$ current:     & $S_{11}(1535)$ & PDG &
$S_{11}(1650)$ & PDG  \\
$M_{R}^{}$ (MeV)                            &
$1540$ & [$1525$--$1545$] &
$1615$   &
[$1645$--$1670$]  \\
$g_{RN\gamma}^{(1)}$      & $0.81$    & & $0.52$   &                  \\
($g_{RN\pi}^{}$, $\lambda$)                 & ($-0.67$, $0.04$)  &
& ($-0.50$, $0.77$)  &                  \\
($g_{RN\eta}^{}$, $\lambda$)                & ($-2.61$, $0.46$)  &
& ($0.98$, $0.91$)   &                  \\
($g_{RN\omega}^{}, f_{RN\omega}^{}$)           & ($38.52$, $-1.39$)  &
& ($86.78$, $6.19$) &                  \\
($g_{RN\rho}^{}, f_{RN\rho}^{}$)               & ($-15.07$, $2.67$) &
& ($-39.01$, $10.84$)         &                  \\
$\Gamma_{R} $ (MeV)                      &
$200$ & [$125$--$175$]  &
$144$   &   [$145$--$185$]   \\
$\gamma_{N\gamma}^{}$ (\%)  &
$0.31$ & [$0.15$--$0.35$] &
$0.24$ & [$0.04$--$0.18$] \\
$\beta_{N\pi}^{}$ (\%)  &
$32$ & [$35$--$55$] &
$27$ & [$60$--$95$] \\
$\beta_{N\eta}^{}$ (\%) &
$65$ & [$45$--$60$] &
$19$ & [$3$--$10$] \\
$\beta_{N\pi\pi}^{}$ (\%) &
$2$ & [$1$--$10$] &
$54$  & [$10$--$20$] \\
\hline
$N_{13}$ current:         & $D_{13}(1520)$ & PDG &
$D_{13}(1700)$ & PDG  \\
$M_{R}^{}$ (MeV)                            &
$\textbf{1520}$ & [$1515$--$1525$] &
$\textbf{1700}$ & [$1650$--$1750$] \\
($g^{(1)}_{RN\gamma}, g^{(2)}_{RN\gamma}$)& ($1.22$, $-0.18$) &
 & ($0.21$, $-0.22$)  &                  \\
$g_{RN\pi}^{}$  & $-1.96$          & &
$0.58$          &                  \\
$g_{RN\eta}^{}$       & $-3.32$   &      &
$-2.40$          \\
($g^{(1)}_{RN\omega}, g^{(2)}_{RN\omega}, g^{(3)}_{RN\omega}$)&
($-289.94$, $242.59$, $\textbf{0.0}$) & &
($341.79$, $-362.81$, $\textbf{0.0}$) & \\
($g^{(1)}_{RN\rho}, g^{(2)}_{RN\rho}, g^{(3)}_{RN\rho}$)   &
($54.58$, $154.53$, $\textbf{0.0}$) & &
($-1.61$, $90.18$, $\textbf{0.0}$)  & \\
$\Gamma_{R} $ (MeV)                      &
$108$ & [$100$--$125$] &
$94$  & [$50$--$150$]  \\
$\gamma_{N\gamma}^{}$ (\%)                  &
$0.41$ & [$0.46$--$0.56$] &
$0.07$ & [$0.01$--$0.05$] \\
$\beta_{N\pi}^{}$ (\%)                      &
$95$ & [$55$--$65$] &
$28$ & [$5$--$15$] \\
$\beta_{N\eta}^{}$ (\%)                     &
$0.03$ & [$0.23 \pm 0.04$] &
$2$ & [$0.0 \pm 1.0$] \\
$\beta_{N\pi\pi}^{}$ (\%)                   &
$5$ & [$40$--$50$] &
$70$ & [$85$--$95$] \\
\hline\hline
\end{tabular}
\endgroup
\end{table*}
%

\begin{figure*}[t!]\centering
\includegraphics[width=\hsize]{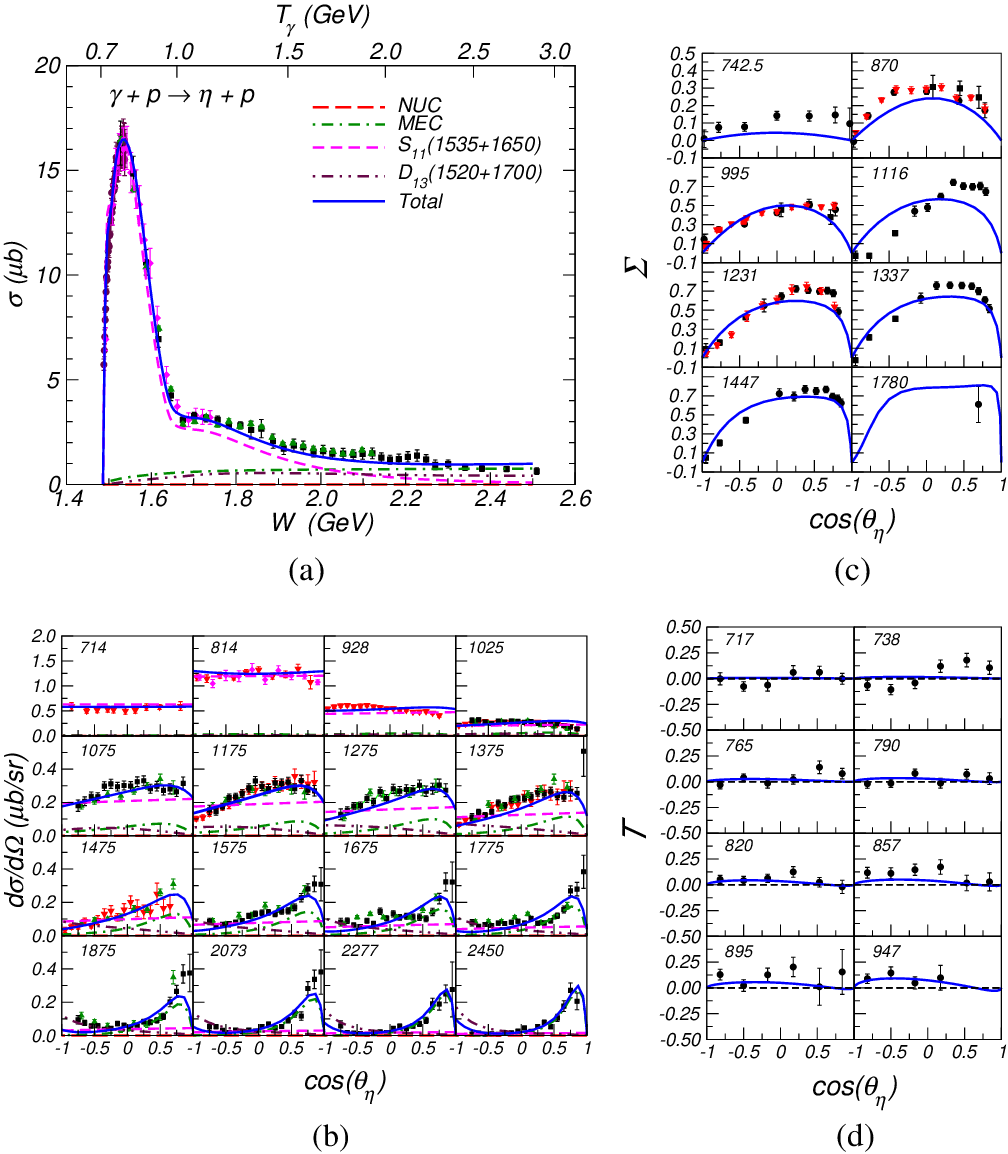}

\caption{(Color online) Results for the reaction $\gamma p \to p \eta$
in model (A), i.e., with the parameters of Table~\ref{tbl:photo4c}.
(a) Total cross section as a function of the total energy of the system $W$.
The line styles identified here apply to all four parts of this
figure.
(b) $\eta$ angular distribution in the center-of-mass frame.
(c) Beam asymmetry $\Sigma$
and (d) target asymmetry $T$ in the center-of-mass frame.
The numbers in (b, c, and d) are the incident photon laboratory energies
$T_\gamma$ in MeV.
In (c and d), only the total results are shown. The data are from
Refs.~\citenum{photo-xsc-data} and \citenum{photo-spin-data}.} \label{fig:photo4c}
\end{figure*}

\subsection{$\bm{\gamma p \to p \eta}$}\label{sec:IIIA}

We first consider the model in which only the $S_{11}(1535)$, $S_{11}(1650)$,
$D_{13}(1520)$, and $D_{13}(1700)$ resonance currents are considered in
addition to the nucleonic and the mesonic currents (cf.\
Fig.~\ref{fig:diagram_photo}). We find that this comprises the minimal set of
resonances that are required to achieve a reasonable description of the
reaction processes listed in Eq.~(\ref{eq:1}). The resulting fitted parameters,
which are obtained by $\chi^2$ fitting to the experimental data shown in 
Fig.~\ref{fig:photo4c} (however, not taking into
account the total cross sections for $\gamma p \to p\eta$)
are given in Table~\ref{tbl:photo4c}. Note that for
photoproduction, in addition to the electromagnetic couplings, only the
hadronic vertices involving the $\eta$ meson are required for the present
calculation. In the $\pi^- p \to n \eta$ reaction, only the hadronic vertices
involving $\eta$ and $\pi$ are needed while all the hadronic vertices given in
the Appendix are required for the calculation of the $NN \to NN\eta$ reaction.

In Table~\ref{tbl:photo4c}, the parameter values in boldface are fixed and
are not allowed to vary during the fitting procedure.
The pure pseudovector coupling choice ($\lambda=0$) at the $NN\eta$ vertex was
motivated by the massless chiral limit of the $\eta$ meson.
(See the Appendix for the definition of $\lambda$.)%
\footnote{We will also consider the pseudoscalar coupling choice
($\lambda=1$). See the discussions in subsection~\ref{sub:NNeta}.}
Also, the most general form of the $RNV$ vertex for spin-3/2 and -5/2
resonances involves three independent coupling constants, as exhibited in
Eqs.~(\ref{VNR32}) and (\ref{VNR52}).
At present, however, information on the corresponding coupling constants is
extremely scarce, especially, on $g^{(3)}_{RNV}$ defined in Eqs.~(\ref{VNR32})
and (\ref{VNR52}).
In the present work, therefore, we allow only two structures at those vertices
by setting the coupling constant $g^{(3)}_{RNV}$ to zero.

In this work, we use the resonance masses from the centroid values quoted in
PDG~\cite{PDG06}. The exceptions to this are the masses of the $S_{11}(1535)$ and
the $S_{11}(1650)$ resonances. They were allowed to vary in the fitting process in
order to reproduce accurately the position of the large cross-section peak
exhibited by the photoproduction data near the threshold. 
The quantities in the square brackets
in Table~\ref{tbl:photo4c} are the range estimates quoted in
PDG~\cite{PDG06} and
are given here for an easy comparison with the fitted values extracted in the
present model. As one can see, some parameter values are considerably outside
the range quoted in PDG. However, we note that, although various parameters are
highly correlated to each other, the data used in the
present study are not sufficient to uniquely constrain the model parameters. As
a result, different parameter sets may provide fits of comparable quality; some
of them are discussed in this paper.

Figure~\ref{fig:photo4c} displays the results for photoproduction observables
corresponding to the parameter set of Table~\ref{tbl:photo4c}. The total cross
section shown in Fig.~\ref{fig:photo4c}(a) provides the line styles used in all
four parts.
We start with the discussion of the $\eta$ angular distribution shown in
Fig.~\ref{fig:photo4c}(b).
As one can see, the flat angular distribution near the threshold is dominated
by the $S_{11}$ resonance.
When the photon incident energy (in the laboratory frame) $T_\gamma$ is larger
than about $1.0$~GeV, both the $D_{13}$ resonance and the mesonic currents
become relevant for reproducing the shape of the measured differential
cross sections.
As the energy increases, the shape becomes more and more forward-peaked, which is
a well-known feature of the $t$-channel mesonic current contribution.
In the energy region of $T_\gamma \approx 1.1$--$2.0$~GeV, some details of
the measured angular distribution are still not well explained, which
indicates that our model should be improved by using a more refined
and quantitative calculation. We leave such an investigation to a future work.
The nucleonic current contribution is negligible because of the very small
coupling constant $g_{NN\eta}^{}$ resulting from the fit,
which is preferred by the small angular distribution measured at backward
angles and higher energies. The nucleonic current contributes mostly at
backward angles and at high energies through the $u$-channel diagram (cf.\
Fig.~\ref{fig:diagram_photo}).
We will come back to this issue later in subsection \ref{sub:NNeta}.

Shown in Fig.~\ref{fig:photo4c}(a) are the total cross sections for $\eta$
photoproduction obtained with the resonance parameters given in
Table~\ref{tbl:photo4c}.
Here, we mention that the total cross section data for photoproduction were
\emph{not\/} included in the global fitting process.
As one can see, the large peak that rises sharply from threshold is due to the
dominating $S_{11}(1535)$ and $S_{11}(1650)$ resonances.
Although both the $D_{13}$ resonance and the mesonic currents are small, their
contribution beyond $W \sim 1.6$ GeV cannot be ignored because of the
interference with the large $S_{11}$ current contribution, a feature 
that has already been pointed out in earlier works~\cite{photo-theory}.
Around $W = 2.0$ GeV, all the resonance and mesonic currents become comparable
to each other, but at higher energies, the mesonic current yields the largest
contribution, and the $S_{11}$ resonance contribution becomes negligible, as
expected from the relative low mass.
The structure shown by the data at $W \sim 2.2$~GeV seems to suggest possible
contributions from other higher mass resonances.
Analyses of this structure and other details are left to a future work.

Figure~\ref{fig:photo4c} also shows the results for  beam and target
asymmetries.
We find that, by and large, the beam asymmetry ($\Sigma$) and the target
asymmetry ($T$) are described reasonably well.%
\footnote{Quite recently, the GRAAL Collaboration \cite{KPBJ08} 
re-analyzed the beam asymmetry in $\gamma p \to p \eta$, revealing a sharp
structure at $W \sim 1.69$ GeV for $\theta_\eta \sim 43^\circ - 103^\circ$ and
suggesting the presence of a narrow resonance.}
The most visible room for improvement exists for the latter, in particular,
at lower energies, where the data show a $\sin(2\theta_\eta)$ dependence, while
the model yields nearly flat and vanishing results. This is a feature common to all 
the parameter sets and not just to the particular parameter set of 
Table~\ref{tbl:photo4c}. The difficulty in reproducing the target asymmetry is a 
known feature from earlier works. (See, in particular, the work of 
Tiator \textit{et al.\/}~\cite{photo-theory}.)
Perhaps, the difficulty in reproducing the target asymmetry may be understood 
if we write the quantities in terms of the four amplitudes $F_i (i=1,...,4)$ a la 
the CGLN decomposition~\cite{CGLN,NL04-NL05}
\begin{equation}
M  = F_1 \,\vec\sigma\cdot\vec\epsilon  
+  i F_2 \,\vec\epsilon\cdot(\hat k \times \hat q)
+ F_3 \,\vec\sigma\cdot\hat k \hat q\cdot\vec\epsilon
+ F_4 \,\vec\sigma\cdot\hat q \hat q\cdot\vec\epsilon .
\label{photo-ampl}
\end{equation}
We then have
\begin{eqnarray}
\frac{d\sigma}{d\Omega} \Sigma &=& \left( |F_2|^2 - |F_3|^2 - |F_4|^2 
- 2 \mbox{Re}[(F_1 + F_3\cos\theta_\eta)F_4^*]\right) \nonumber \\
&& \mbox{} \times\sin^2 \theta_\eta , \nonumber \\
\frac{d\sigma}{d\Omega} T &=& 2\mbox{Im}
\left[(-F_2 + F_3 + F_4\cos\theta_\eta) F_1^* \right. \nonumber \\
&& \mbox{} \left. + (F_3 + F_4\cos\theta_\eta) F_4^*\sin^2\theta_\eta 
\right]\sin\theta_\eta,  \nonumber \\
\frac{d\sigma}{d\Omega} P &=& -2\mbox{Im}
\left[(F_2 + F_3 + F_4\cos\theta_\eta) F_1^* \right. \nonumber \\
&& \mbox{} \left. + (F_3 + F_4\cos\theta_\eta) F_4^*\sin^2\theta_\eta 
\right]\sin\theta_\eta ,
\label{spin-asym}
\end{eqnarray}
where the expression for the recoil polarization $P$  is also given.
The above results reveal that, unlike the beam asymmetry,  the target asymmetry 
involves the imaginary part of the product of amplitudes $F_i$. 
This means that this observable is more sensitive to the effects of the 
final state interaction, a feature that is not accounted 
for explicitly in the present type three-level calculations. 
Furthermore, it is also clear that the same difficulty should  be 
present in describing the recoil polarization.
In fact, this seems to be the case, as reported in Ref.~\citenum{Merkel07}.

\begin{figure*}[t!]\centering
\includegraphics[width=\hsize]{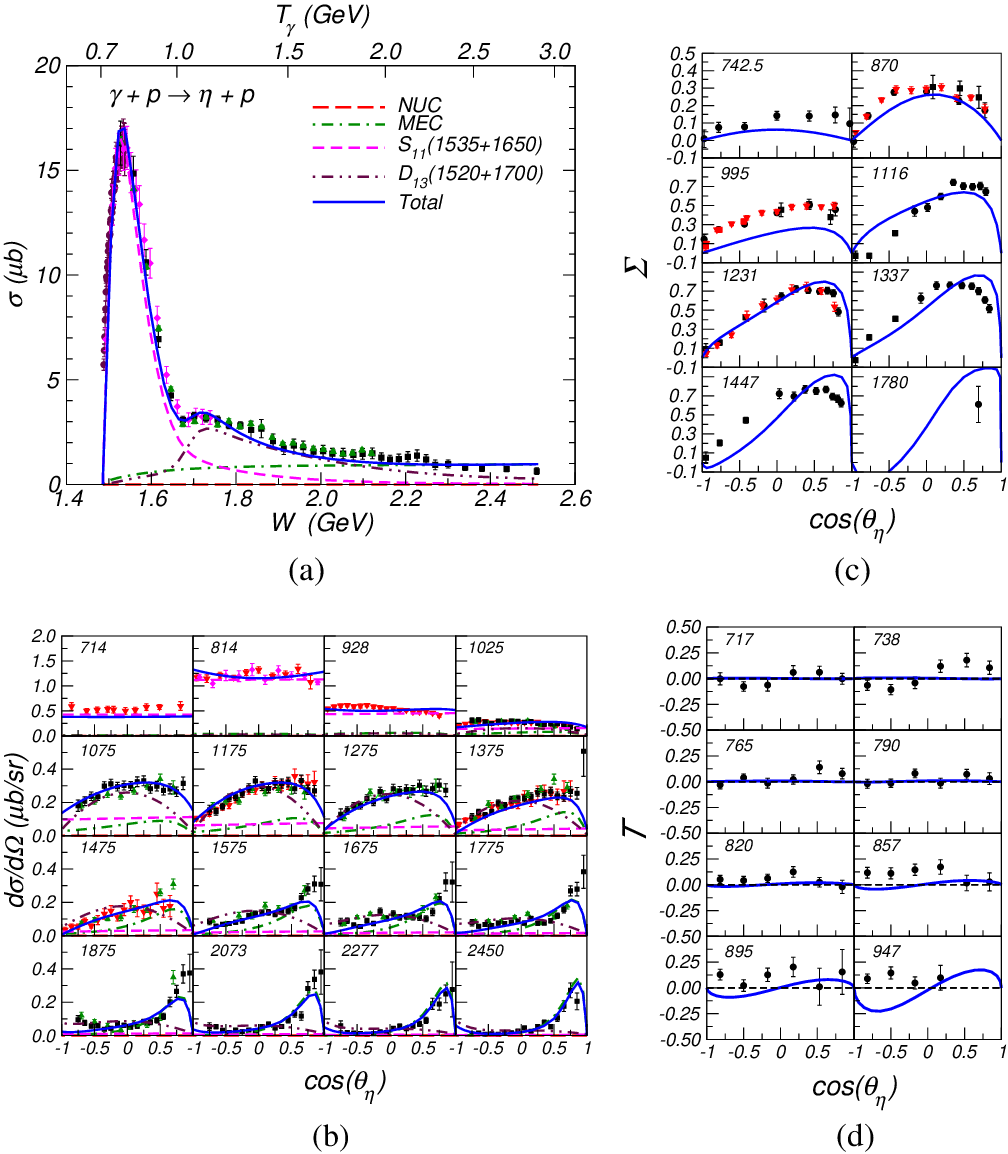}

\caption{(Color online)
Same as Fig.~\ref{fig:photo4c} but for model (B), i.e.,
with the fitted parameter set of Table~\ref{tbl:photo4d}.}
\label{fig:photo4d}
\end{figure*}

\begin{figure*}[t!]\centering
\includegraphics[width=\hsize]{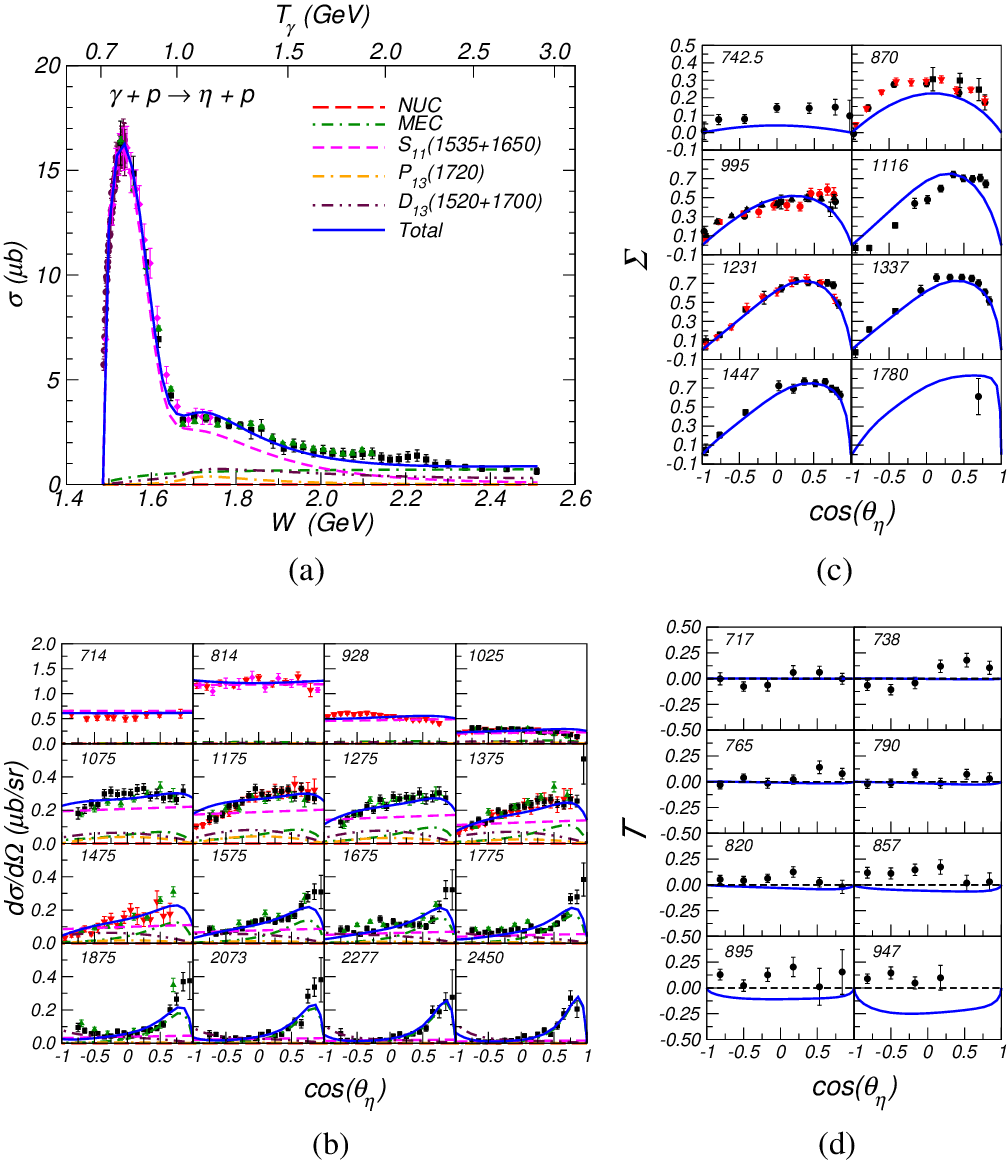}
\caption{(Color online)
Same as Fig.~\ref{fig:photo4c}, but for model (C), i.e.,
with the parameter set of Table~\ref{tbl:photo5a}.}
\label{fig:photo5a}
\end{figure*}

\begin{figure*}[t!]\centering
\includegraphics[width=\hsize]{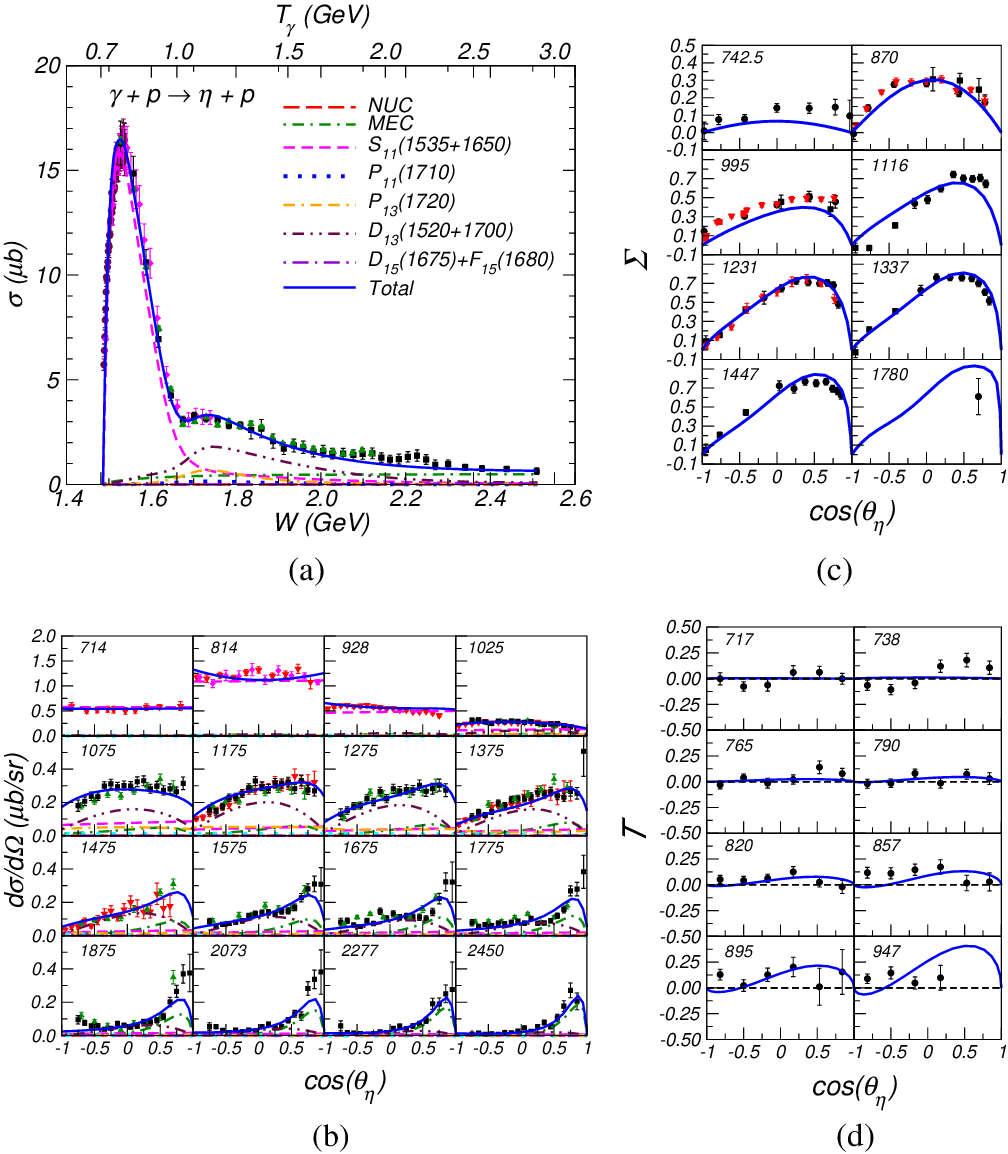}

\caption{(Color online)
Same as Fig.~\ref{fig:photo4c}, but for model (D), i.e.,
with the parameter set of Table~\ref{tbl:photo52b}.}
\label{fig:photo52b}
\end{figure*}

We anticipate here that, in contrast to low energies, the target asymmetry becomes 
sensitive to the dynamical content of the model as the energy increases (cf. the results in
Figs.~\ref{fig:photo4c}, \ref{fig:photo4d}, \ref{fig:photo5a}, and \ref{fig:photo52b}). Therefore, we
expect that this observable at higher energies will be useful in imposing extra constraints on 
the model parameters.

\begin{table*}[t!]\centering
\caption{\label{tbl:photo4d}
Another set of fitted parameters [model (B)]. See the caption of
Table~\ref{tbl:photo4c} for details.}
\begingroup 
\begin{tabular}{l@{\qquad}r@{\qquad}l@{\qquad}r@{\qquad}l}
\hline\hline
Nucleonic current: & & & & \\
($g_{NN\eta}^{}$, $\lambda$)        &
($0.088$, $\textbf{0.0}$)  &  &      &           \\
\hline
Mesonic current: & & & &  \\
$\Lambda^*_v$ (MeV)                      &  $1202$
&                &                  &                  \\
\hline
$N_{11}$ current:                        &
$S_{11}(1535)$ & PDG & $S_{11}(1650)$ & PDG \\
$M_{R}^{}$ (MeV)                            &
$1527$ & [$1525$--$1545$] &
$1625$ & [$1645$--$1670$] \\
$g_{RN\gamma}^{(1)}$  & $-0.55$  & & $0.38$    & \\
($g_{RN\pi}^{}$, $\lambda$) &  ($0.43$, $0.005$)  &  &
($-0.37$, $0.84$)  &                  \\
($g_{RN\eta}^{}$, $\lambda$)  & ($1.92$, $0.44$)   & &
($0.68$, $0.86$)   \\
($g_{RN\omega}^{}, f_{RN\omega}^{}$)  & ($-3.86$, $0.79$) & &
($8.81$, $-29.85$)  \\
($g_{RN\rho}^{}, f_{RN\rho}^{}$)  & ($29.18$, $8.40$) & &
($4.09$, $3.63$)  \\
$\Gamma_{R} $ (MeV)   &
$99$ & [$125$--$175$]  &
$162$ & [$145$--$185$] \\
$\gamma_{N\gamma}^{}$ (\%)   &
$0.27$ & [$0.15$--$0.35$] &
$0.12$ & [$0.04$--$0.18$] \\
$\beta_{N\pi}^{}$ (\%)    &
$26$ & [$35$--$55$]   &
$13$ & [$60$--$95$] \\
$\beta_{N\eta}^{}$ (\%)                     &
$63$ & [$45$--$60$]       &
$9$ & [$3$--$10$]     \\
$\beta_{N\pi\pi}^{}$ (\%)                   &
$10$ & [$1$--$10$] &
$78$ & [$10$--$20$] \\
\hline
$N_{13}$ current:                        &
$D_{13}(1520)$ & PDG & $D_{13}(1700)$ &  PDG  \\
$M_{R}^{}$ (MeV)                            &
$\textbf{1520}$ & [$1515$--$1525$] &
$\textbf{1700}$ & [$1650$--$1750$] \\
($g^{(1)}_{RN\gamma}, g^{(2)}_{RN\gamma}$) &
($-0.025$, $-2.51$) &  &
($0.59$, $-0.25$)  &   \\
$g_{RN\pi}^{}$  & $1.96$ & &
$0.003$     &    \\
$g_{RN\eta}^{}$  & $2.81$   &        & $-7.17$ & \\
($g^{(1)}_{RN\omega}, g^{(2)}_{RN\omega}, g^{(3)}_{RN\omega}$) &
($-34.75$, $2.91$, $\textbf{0.0}$) & &
($-10.55$, $14.65$, $\textbf{0.0}$) & \\
($g^{(1)}_{RN\rho}, g^{(2)}_{RN\rho}, g^{(3)}_{RN\rho}$)   &
($-29.66$, $24.37$, $\textbf{0.0}$) &  &
($-6.33$, $-2.39$, $\textbf{0.0}$) &  \\
$\Gamma_{R} $ (MeV)    &
$109$ & [$100$--$125$] &
$119$ & [$50$--$150$] \\
$\gamma_{N\gamma}^{}$ (\%)                  &
$0.70$ & [$0.46$--$0.56$] &
$0.24$ & [$0.01$--$0.05$]  \\
$\beta_{N\pi}^{}$ (\%)                      &
$95$ & [$55$--$65$]  &
$0.0006$ & [$5$--$15$]     \\
$\beta_{N\eta}^{}$ (\%)                     &
$4$ & [$0.23 \pm 0.04$]  &
$12$ & [$0.0 \pm 1.0$]    \\
$\beta_{N\pi\pi}^{}$ (\%)                   &
$0.02$ & [$40$--$50$]  &
$88$ & [$85$--$95$] \\
\hline\hline
\end{tabular}
\endgroup
\end{table*}
%

Another parameter set resulting from the global fit, employing the same 
set of nucleon resonances as in Table~\ref{tbl:photo4c}, is shown in
Table~\ref{tbl:photo4d}. 
This set [model (B)] was obtained by using 
different starting values for the search in parameter space during the 
fit procedure. As a result,
many parameter values of this set are quite different from those of
Table~\ref{tbl:photo4c} not only in magnitude but also in relative
signs for some coupling constants, which provides an indication of the general
reliability of such global fits. The corresponding observables are shown in
Fig.~\ref{fig:photo4d}. Here, although both the measured total and differential
cross sections are reproduced with a comparable fit quality to the results in
Fig.~\ref{fig:photo4c}, the dynamical content is quite different. In
particular, the $D_{13}(1700)$ resonance contribution dominates over the other
currents in the energy region of $T_\gamma\approx 1.05$--$1.5$~GeV (i.e.,
$W\approx 1.7$--$1.9$~GeV).
It is interesting to note that, searching for the pole positions of the $T$-matrix
in the complex plane, as well as performing Breit--Wigner parameterizations of
resonances, the recent partial-wave analysis by Arndt \textit{et
al.}~\cite{ABSW06} of the $\pi^{\pm}p$ elastic and charge-exchange processes,
combined with the reaction $\pi^- p \to n \eta$, finds no $D_{13}(1700)$
resonance. Also, some of the coupled-channels dynamical models
\cite{KHKS00,GHHS03} find no necessity for this resonance to fit the $\pi N$
phase shifts. In this respect, the present results corresponding to the
parameter set of Table~\ref{tbl:photo4c}, where the contribution of the
$D_{13}(1700)$ resonance is much smaller than that of Table~\ref{tbl:photo4d},
are more in line with these findings. It would be most interesting to see
if the inclusion of the $\gamma N$ channel in those coupled-channels analyses
mentioned above would require the $D_{13}(1700)$ resonance.
Overall, the spin observables in Fig.~\ref{fig:photo4d} exhibit the same
features as in Fig.~\ref{fig:photo4c}, but for higher energies, the target
asymmetry shows very different angular dependences for the two parameter sets,
which points to the importance of spin asymmetries in reducing the ambiguities
that otherwise would exist.

\begin{table*}[t!]\centering
\caption{\label{tbl:photo5a}
The parameter set for model (C), where
the $P_{13}(1720)$ resonance is added
to see whether it further improves the fit. No $NN \to NN\eta$ data were
used for this fit.}
\begingroup 
\begin{tabular}{l@{\qquad}r@{\qquad}l@{\qquad}r@{\qquad}l@{\qquad}r@{\qquad}l}
\hline\hline
Nucleonic current: & & & & \\
($g_{NN\eta}^{}$, $\lambda$)   &
($0.003$, $\textbf{0.0}$)  &   &  &  \\
\hline
Mesonic current: & & & &  \\
$\Lambda^*_v$ (MeV) &  $1162$ &   &   &   \\
\hline
$N_{11}$ current:                        &
$S_{11}(1535)$ & PDG & $S_{11}(1650)$ & PDG  \\
$M_{R}^{}$ (MeV)        &
$1539$ & [$1525$--$1545$] &
$1617$ & [$1645$--$1670$] \\
$g_{RN\gamma}^{(1)}$   & $1.01$ &  & $0.58$    &  \\
($g_{RN\pi}^{}$, $\lambda$)  &
($0.64$, $0.008$)  &  &
($0.42$, $0.28$)   &  \\
($g_{RN\eta}^{}$, $\lambda$) &
($2.70$, $0.36$)   &  &
($-1.11$, $0.62$)  &  \\
$\Gamma_{R} $ (MeV)  &
$200$ & [$125$--$175$]  &
$144$ & [$145$--$185$] \\
$\gamma_{N\gamma}^{}$ (\%)   &
$0.28$ & [$0.15$--$0.35$]  &
$0.16$ & [$0.04$--$0.18$]  \\
$\beta_{N\pi}^{}$ (\%)     &
$30$ & [$35$--$55$]  &
$19$ & [$60$--$95$] \\
$\beta_{N\eta}^{}$ (\%)   &
$69$ & [$45$--$60$] &
$55$ & [$3$--$10$]  \\
$\beta_{N\pi\pi}^{}$ (\%)                   &
$0.72$ & [$1$--$10$]  &
$25$ & [$10$--$20$]  \\
\hline
$N_{13}$ current:     &
$D_{13}(1520)$ & PDG & $D_{13}(1700)$ & PDG & $P_{13}(1720)$ & PDG \\
$M_{R}^{}$ (MeV)                            &
$\textbf{1520}$ & [$1515$--$1525$] &
$\textbf{1700}$ & [$1650$--$1750$] &
$\textbf{1720}$ & [$1700$--$1750$]  \\
($g^{(1)}_{RN\gamma}, g^{(2)}_{RN\gamma}$) &
($1.45$, $-0.14$)  &  &
($0.01$, $0.48$)   &  &
($0.42$, $1.44$)   &  \\
$g_{RN\pi}^{}$   &
$-2.01$ &  & $0.84$ &   & $0.42$  & \\
$g_{RN\eta}^{}$  & $-2.60$  &  & $-2.64$ &   & $0.74$ &  \\
$\Gamma_{R} $ (MeV)                      &
$114$ & [$100$--$125$] &
$126$ & [$50$--$150$]  &
$192$ & [$150$--$300$]  \\
$\gamma_{N\gamma}^{}$ (\%)                  &
$0.005$ & [$0.46$--$0.56$] &
$0.05$  & [$0.01$--$0.05$] &
$0.12$  & [$0.003$--$0.01$] \\
$\beta_{N\pi}^{}$ (\%)                      &
$95$ & [$55$--$65$] &
$44$ & [$5$--$15$]  &
$95$ & [$10$--$20$]  \\
$\beta_{N\eta}^{}$ (\%)                     &
$4$  & [$0.23 \pm 0.04$]  &
$1.55$  & [$0.0 \pm 1.0$] &
$2.69$  & [$4.0 \pm 1.0$] \\
$\beta_{N\pi\pi}^{}$ (\%)                   &
$0.02$ & [$40$--$50$]  &
$55$   & [$85$--$95$]  &
$2.19$   &  [$>70$]  \\
\hline\hline
\end{tabular}
\endgroup
\end{table*}
%

Table~\ref{tbl:photo5a} [model (C)] displays a parameter set including the $P_{13}(1720)$
resonance, in addition to those considered in the previous two sets.
Here, we have considered only the $\gamma p \to p \eta$ and the $\pi^- p \to n \eta$
reaction data in the fitting procedure.
The corresponding results for the observables are shown in
Fig.~\ref{fig:photo5a}.
As can be seen, the inclusion of the $P_{13}(1720)$ resonance does not improve
significantly the description of the data for this photon-induced reaction.
However, as we will see in the following subsection, this resonance
considerably improves the fit quality of the hadronic $\pi^- p \to n \eta$
reaction at higher energies.

\begin{table*}[t!]\centering
\caption{Same as Table~\ref{tbl:photo4c}.
The parameter set for model (D). Here,
more resonances are added here to see
whether they would further improve the fit. Here we have considered only the
$\gamma p \to p \eta$ reaction in the fitting procedure.}
\begin{tabular}{l@{\qquad}r@{\qquad}l@{\qquad}r@{\qquad}l@{\qquad}r@{\qquad}l}
\hline\hline
Nucleonic current: & & & & \\
($g_{NN\eta}^{}$, $\lambda$)                &
($0.07$, $\textbf{0.0}$)   &      &      &           \\
\hline
Mesonic current: & & & &  \\
$\Lambda^*_v$ (MeV)                      &  $1113$
&                &                  &                  \\
\hline
$N_{11}$ current:                        &
$S_{11}(1535)$ & PDG & $S_{11}(1650)$ & PDG & $P_{11}(1710)$  & PDG  \\
$M_{R}^{}$ (MeV)                            &
$1521$ & [$1525$--$1545$] &
$1632$ & [$1645$--$1670$] &
$\textbf{1710}$ & [$1680$--$1740$] \\
($g_{RN\gamma}^{(1)}g_{RN\eta}, \lambda)$  &
($1.22$, $0.33$)  & &
($-0.49$, $1.00$)  & &
($-0.63$, $1.00$)    &    \\
$\Gamma_{R} $ (MeV)       &
$110$ & [$125$--$175$] &
$174$ & [$145$--$185$] &
$250$ & [$50$--$250$]  \\
$\gamma_{N\gamma}^{}$ (\%)                  &
$\textbf{0.26}$ & [$0.15$--$0.35$] &
$\textbf{0.06}$ & [$0.04$--$0.18$] &
$\textbf{0.01}$ & [$0.002$--$0.05$] \\
$\beta_{N\pi}^{}$ (\%)                      &
$30$ & [$35$--$55$] &
$18$ & [$60$--$95$] &
$81$   & [$10$--$20$] \\
$\beta_{N\eta}^{}$ (\%)                     &
$65$ & [$45$--$60$] &
$26$ & [$3$--$10$] &
$19$ & [$6.2 \pm 1.0$] \\
$\beta_{N\pi\pi}^{}$ (\%)         &
$5$ & [$1$--$10$] &
$56$ & [$10$--$20$] &
$0.3$ & [$40$--$90$] \\
\hline
$N_{13}$ current:           &
$P_{13}(1720)$ & PDG & $D_{13}(1520)$ & PDG & $D_{13}(1700)$ & PDG \\
$M_{R}^{}$ (MeV)                            &
$\textbf{1720}$ & [$1700$--$1750$] &
$\textbf{1520}$ & [$1515$--$1525$] &
$\textbf{1700}$ & [$1650$--$1750$] \\
$g^{(1)}_{RN\gamma}g_{RN\eta}^{}$         & $-0.85$          &
$0.22$          &  $0.37$            &                \\
$g^{(2)}_{RN\gamma}g_{RN\eta}^{}$         &  $1.60$          &
$-10.32$         &  $-2.89$           &                \\
$\Gamma_{R} $ (MeV)                      &
$184$ & [$150$--$300$] &
$136$ & [$100$--$125$] &
$135$ & [$50$--$150$] \\
$\gamma_{N\gamma}^{}$ (\%)                  &
$\textbf{0.12}$ & [$0.003$--$0.01$] &
$\textbf{0.10}$ & [$0.46$--$0.56$] &
$\textbf{0.54}$ & [$0.01$--$0.05$] \\
$\beta_{N\pi}^{}$ (\%)                      &
$95$ & [$10$--$20$] &
$76$ & [$55$--$65$] &
$60$ & [$5$--$15$] \\
$\beta_{N\eta}^{}$ (\%)                     &
$5$   & [$4.0 \pm 1.0$]  &
$0.04$ & [$0.23 \pm 0.04$] &
$4$ & [$0.0 \pm 1.0$] \\
$\beta_{N\pi\pi}^{}$ (\%)                   &
$0.04$ & [$>70$] &
$24$         & [$40$--$50$] &
$36$ & [$85$--$95$] \\
\hline
$N_{15}$ current:                        &
$D_{15}(1675)$ & PDG & $F_{15}(1680)$ &    PDG &  \\
$M_{R}^{}$ (MeV)                            &
$\textbf{1675}$ & [$1670$--$1680$] &
$\textbf{1680}$ & [$1680$--$1690$] \\
$g^{(1)}_{RN\gamma}g_{RN\eta}$  &
$3.77$  &       &  $0.44$ \\
$g^{(2)}_{RN\gamma}g_{RN\eta}$         &
$14.12$   &     & $-0.90$ &                   \\
$\Gamma_{R} $ (MeV)                      &
$171$ & [$130$--$165$] &
$139$ & [$120$--$140$] \\
$\gamma_{N\gamma}^{} $  (\%)               &
$\textbf{0.02}$  & [$0.004$--$0.022$] &
$\textbf{0.25}$  & [$0.21$--$0.32$] \\
$\beta_{N\pi}^{} $      (\%)               &
$\textbf{45}$ & [$35$--$45$] &
$\textbf{65}$ & [$65$--$70$] \\
$\beta_{N\eta}^{} $      (\%)               &
$2$ & [$0.0 \pm 1.0$] &
$0.0001$ & [$0.0 \pm 1.0$] \\
$\beta_{N\pi\pi}^{}$     (\%)               &
$53$ & [$50$--$60$] &
$35$ & [$30$--$40$] \\
\hline\hline
\end{tabular}
\label{tbl:photo52b}
\end{table*}
%

Next, we extend our model by including all of the well-established resonances
in the mass region of $1500 \sim 1700$~MeV in order to verify whether these
resonances can lead to a qualitatively superior description of the data
compared to the previous cases, where a more limited set of resonances was
considered. To this end, we concentrate only on the $\gamma p \to p \eta$
reaction. Table~\ref{tbl:photo52b} shows the parameter set, where the
$D_{15}(1675)$ and $F_{15}(1680)$, as well as the $P_{11}(1710)$, resonances are
included, in addition to those considered in Table~\ref{tbl:photo5a}. Here,
following Ref.~\citenum{NH04-NH05}, we treat the branching ratio $\beta_{N\pi}^{}$ as a
free parameter to be fitted while the branching ratio $\beta_{N\eta}^{}$ is
extracted from the product of the coupling constants
$g_{RN\eta}^{}g_{RN\gamma}^{}$ in conjunction with the assumed branching ratio
$\gamma_{N\gamma}^{}$ for the radiative decay. The resulting observables with
the parameter set of Table~\ref{tbl:photo52b} [model (D)] are shown in
Fig.~\ref{fig:photo52b}. We see that the overall fit quality does not change
significantly from the fit qualities for the previous sets. Here, the $D_{13}(1700)$ resonance gives
the largest contribution to the cross section in the energy region of $W =
1.7$--$2.0$~GeV, a feature similar to that  already exhibited in
Fig.~\ref{fig:photo4d}.

%
%
\begin{figure}[t!]\centering
\includegraphics[width=7.5 cm]{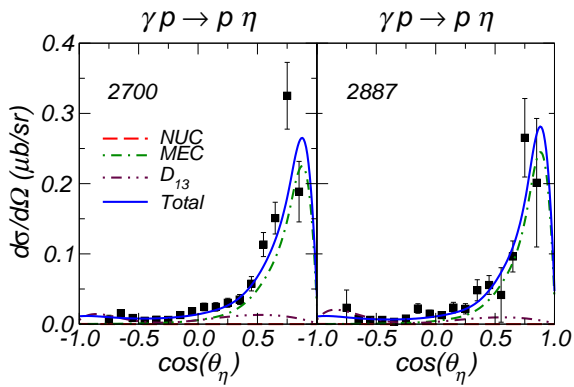} \\
\vspace{0.3 cm}
\includegraphics[width=7.5 cm]{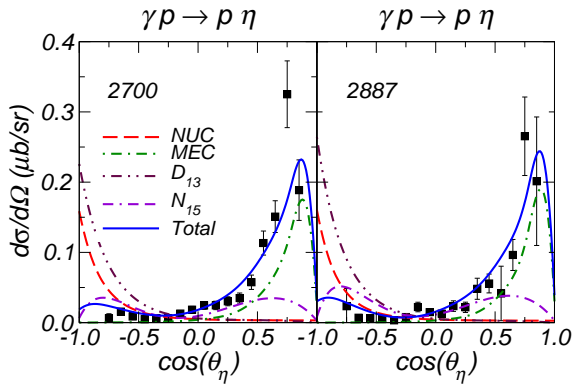} \\
\vspace{0.3 cm}
\includegraphics[width=7.5 cm]{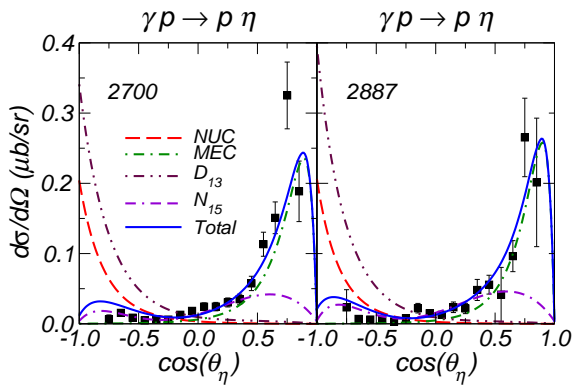}
\caption{(Color online)
$\eta$-meson angular distributions in the center-of-mass frame in
$\gamma p \to p\eta$ at $T_\gamma = 2700$~MeV and $2887$~MeV.
(a) The results corresponding to the parameter set of Table~\ref{tbl:photo52b},
with $g_{NN\eta}=0.07$.
(b,c) The results obtained with two other parameter sets (not given in this
work) using the pseudovector ($\lambda=0$) and the pseudoscalar ($\lambda=1$)
coupling choices, respectively, at the $NN\eta$ vertex.
The resulting $NN\eta$ coupling constant values are $g_{NN\eta}=1.62$
and $1.38$, respectively.
The contributions from the other nucleon resonances are practically negligible
at these energies and are not displayed.
$N_{15}$ is the sum of spin-$5/2$ resonance contributions.
The data are from Cred{\'e} \textit{et al.\/}~\cite{photo-xsc-data}.}
\label{fig:photo52b_he}
\end{figure}
%
%

%
\begin{figure}[t!]\centering
\includegraphics[width=75mm,clip=]{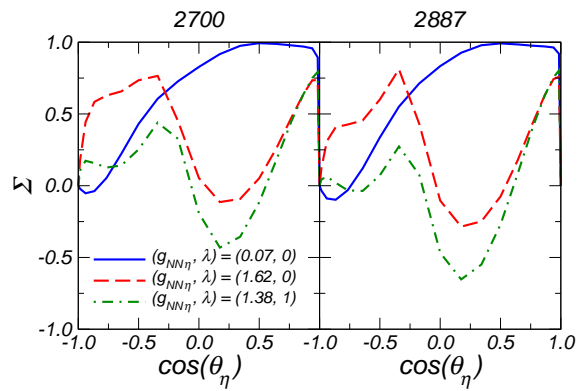}\\
\vspace{.3 cm}
\includegraphics[width=75mm,clip=]{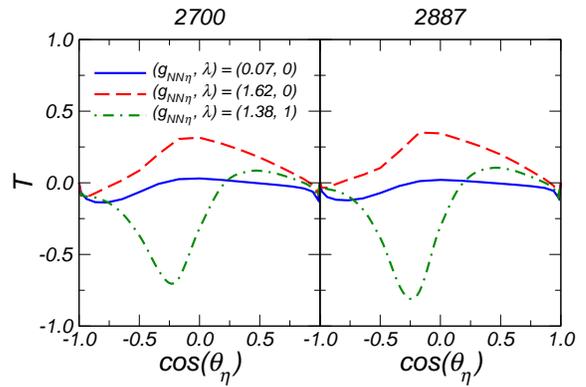}
\caption{(Color online) %
(a) Beam asymmetry $\Sigma$ and (b) target asymmetry $T$ in
$\gamma p \to p\eta$ at $T_\gamma = 2700$~MeV and $2887$~MeV as
functions of $\eta$-emission angle in the center-of-mass frame.
The solid curves represent the results obtained with the parameter set of
Table~\ref{tbl:photo52b}, which correspond to Fig.~\ref{fig:photo52b_he}(a).
The dashed and the dash-dotted curves are obtained with the two other parameter
sets that correspond to Figs.~\ref{fig:photo52b_he}(b) and (c), respectively.}
\label{fig:photo52b_he_sb}
\end{figure}
%
%

\subsection{$NN\eta$ Coupling Constant}
\label{sub:NNeta}

In the previous subsection, we have shown that the present calculations yield
very small values of the coupling constant $g_{NN\eta}^{}$
--- compatible with zero --- due to the smallness of the measured cross
sections at backward angles where $T_\gamma$ is large. As mentioned before, the
$\eta$ angular distribution becomes very sensitive to $g_{NN\eta}^{}$ at these
kinematics through the $u$-channel nucleonic current contribution.
However, one must be cautious in drawing conclusions about the extracted
value of $g_{NN\eta}^{}$ from calculations based on approaches such as the
present one.
This is due to the fact that we cannot completely discard the possibility of a
relatively large nucleonic current contribution interfering destructively with
contributions from resonances in order to yield the $\eta$ angular distributions
observed at those kinematics. This point is illustrated in
Fig.~\ref{fig:photo52b_he}, where the results for the $\eta$ angular distribution
are shown, together with the data reported by Cred\'e \textit{et al.\/}
(CB/ELSA Collaboration)~\cite{photo-xsc-data}, at the two highest energies.
Figure~\ref{fig:photo52b_he}(a) shows the results corresponding to the
parameter set of Table~\ref{tbl:photo52b}, where the nucleonic current
contribution is practically zero%
\footnote{Note that the value of $g_{NN\eta}^{}$ is $0.07$ in 
Table~\ref{tbl:photo52b}.}
and cannot be seen in the figure. The corresponding nucleon resonance
contributions are also very small. The mesonic current dictates to a large
extent the behavior of the angular distribution at forward angles.
Figures~\ref{fig:photo52b_he}(b) and (c) display the results corresponding to two
additional parameter sets (not given here) with the same set of nucleon
resonances as in Table~\ref{tbl:photo52b}. Overall, the two parameter sets
yield a fit quality comparable to that of Fig.~\ref{fig:photo52b}, but with
very different resonance parameter values, which points to the ambiguity of the
fit results if one relies solely on differential cross-section data. Shown in
Fig.~\ref{fig:photo52b_he}(b) is the result obtained by using pure pseudovector
coupling ($\lambda=0$) as in Fig.~\ref{fig:photo52b} while in
Fig.~\ref{fig:photo52b_he}(c), pure pseudoscalar coupling ($\lambda=1$) is
adopted. The corresponding $NN\eta$ coupling constants are $g_{NN\eta}^{}=1.62$
and $1.38$, respectively. As can be seen in Figs.~\ref{fig:photo52b_he}(b) and (c),
in these cases the nucleonic current contribution at backward angles is as
large as that of the mesonic current at forward angles. However, the $D_{13}$
resonance contribution exhibits an angular dependence similar to that of the nucleonic
current with a comparable magnitude and interferes \emph{destructively\/} with
the nucleonic current. The destructive interference is almost complete and
results in very small cross sections at backward angles as observed in the
data. Overall, everything else being very similar between
Figs.~\ref{fig:photo52b_he}(b) and (c), we find no real sensitivity of the
differential cross sections at these energies as to whether 
pseudovector or pseudoscalar couplings are employed.

The situation changes when one considers spin observables.
While there is no real difference at lower energies, at the high
energies ($T_\gamma =2700 \sim 2887$~MeV) considered here, the beam asymmetry
$\Sigma$ shown in Fig.~\ref{fig:photo52b_he_sb}(a) can distinguish clearly
between the parameter sets corresponding to Fig.~\ref{fig:photo52b_he}(a) on
the one hand and Figs.~\ref{fig:photo52b_he}(b) and (c) on the other.
The marked differences are due to the marked differences in the values for the
$NN\eta$ coupling constant $g_{NN\eta}^{}$, which is vanishingly small
($g_{NN\eta}^{}=0.07$) for the set corresponding to
Fig.~\ref{fig:photo52b_he}(a) and much larger (and about the same,
$g_{NN\eta}^{}=1.62$ and $g_{NN\eta}^{}=1.38$, respectively) for
Figs.~\ref{fig:photo52b_he}(b) and (c).
This finding shows that the beam asymmetry at higher energies can impose more
stringent constraints, in particular, on the $NN\eta$ coupling constant.
In any case, judging from the results in the present investigation, we expect
the upper limit of the $NN\eta$ coupling constant to be not much larger than
$g_{NN\eta}^{}\approx 1.7$.

These parameter sets also lead to noticeable differences --- albeit not
quite as large --- for the target asymmetry $T$, as shown in
Fig.~\ref{fig:photo52b_he_sb}(b).
Of particular importance, however, is that this observable may
distinguish between the use of the pseudoscalar or the pseudovector coupling
at the $NN\eta$ vertex, which the results of Figs.~\ref{fig:photo52b_he}(b) and (c)
and the respective curves of Fig.~\ref{fig:photo52b_he_sb}(a) cannot do. 
Of course, one should keep in mind that the target asymmetry can be more 
sensitive to the effects of the FSI than the beam asymmetry does,
as Eq.(\ref{spin-asym}) indicates. Therefore, one should be 
cautious in drawing strong conclusions from the present results, which 
do not account for the FSI explicitly.

Before closing this subsection, we remark that, quite recently, the authors of
Ref.~\citenum{FMU06}  addressed the issue of chiral symmetry in $\eta$-meson
photoproduction through the pseudoscalar-pseudovector mixing parameter
$\lambda$ at the $NN\eta$ vertex by investigating this reaction close to the
threshold.
Our study reveals that one must be careful with such an investigation for the
reasons mentioned above. In particular, if the $NN\eta$ coupling constant
turns out to be very small, it will be very difficult to determine the value
of the mixing parameter $\lambda$.

We also note that in Ref.~\citenum{TBK94-2} an alternative way of extracting the 
$NN\eta$ coupling is discussed, where, in contrast to the present work, one makes 
use of the cross section data at very low energies. There, the assumption is made 
that all the $\eta$ production mechanisms are known, except for the nucleonic current. 
The extracted value of $g_{NN\eta}^{}$ is consistent with the present findings.

\subsection{$\bm{\pi^- p \to n \eta}$}

\begin{figure*}[t!]\centering
\includegraphics[width=\hsize]{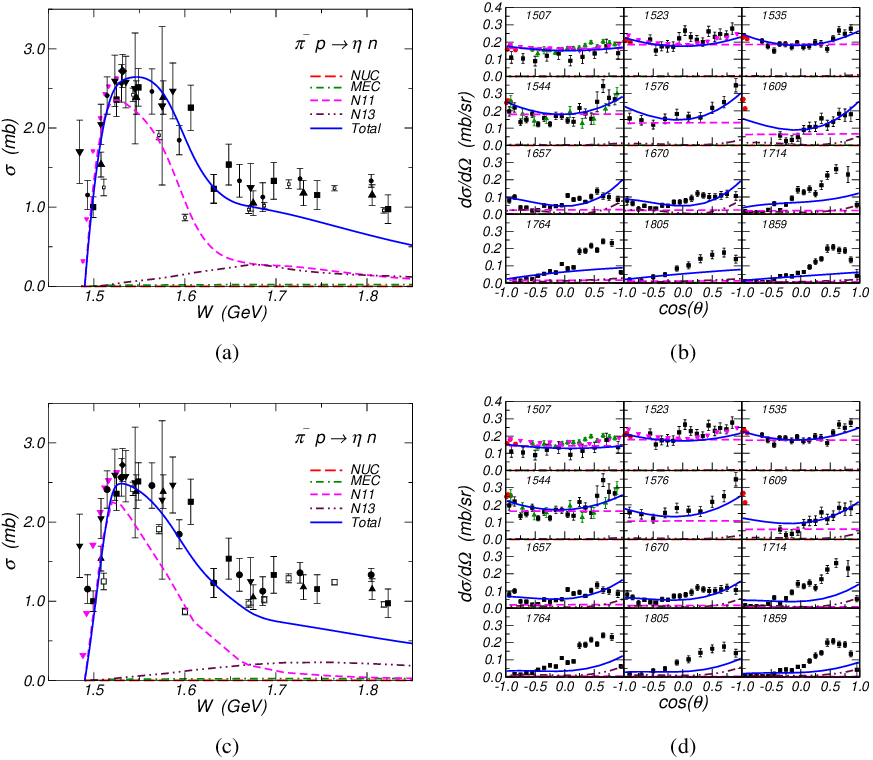}

\caption{\label{fig:pipetan4cd}%
(Color online)
Results for $\pi^- p \to n \eta$ corresponding to the parameter set of
Table~\ref{tbl:photo4c}.
(a) Total cross section as a function of the total energy of the system $W$.
(b) $\eta$ angular distribution in the center-of-mass frame.
Here, $N11$ stands for $S_{11}(1535)+S_{11}(1650)$ contributions and
$N13=D_{13}(1520)+D_{13}(1700)$ contributions.
(c and d) Same as (a) and (b), but with the parameter set of
Table~\ref{tbl:photo4d}.
The numbers in (b) and (d) denote the total center-of-mass energy $W$ in MeV.
The labelings of the curves in (b and d) are the same to that of (a and c).
The experimental data are from Ref.~\citenum{piNetaN-data,CB05}.}
\end{figure*}

In this subsection, we discuss the $\pi^- p \to n \eta$ reaction with the
parameter sets determined above.
The results for the total and the differential cross sections for the reaction
$\pi^- p \to n \eta$ corresponding to the parameter sets of
Tables~\ref{tbl:photo4c} and \ref{tbl:photo4d} are displayed in
Figs.~\ref{fig:pipetan4cd}(a) and (b) and in Figs.~\ref{fig:pipetan4cd}(c) and (d),
respectively.
The cross-section results show that the various dynamical contributions of
the two sets are very similar.
The total cross section is rather well reproduced up to $W\approx 1.6$~GeV,
where it is dominated by the $S_{11}$ resonances, especially, by the
$S_{11}(1535)$ resonance.
Here, both the nucleonic and the mesonic currents yield very small contributions.
However, we do not reproduce the total cross section at higher energies due to
the absence of the higher-mass resonances in these parameter sets and the
absence of the $N\pi\pi$ contribution via the coupled
channel~\cite{IOV01,GHHS03} in this model.
For differential cross sections, we again note that these parameter sets
are unable to reproduce the structure exhibited by the data at higher energies.

\begin{figure*}[t!]
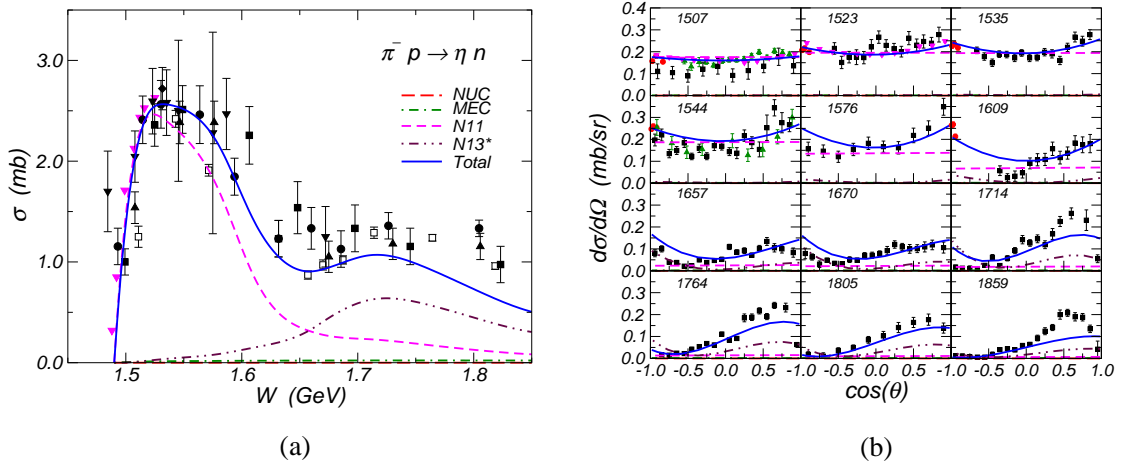
\centering
\includegraphics[width=7.0 cm]{fig11a.eps}
\hspace{0.5 cm}
\includegraphics[width=7.0 cm]{fig11b.eps}
\caption{\label{fig:pipetan5a}%
(Color online)
Same as Fig.~\ref{fig:pipetan4cd} but with  the parameter set
shown in Table~\ref{tbl:photo5a}.
Here, $N13^*$ is the sum of contributions from
$D_{13}(1520)$, $D_{13}(1700)$, and $P_{13}(1720)$.}
\end{figure*}

As we have shown before, the inclusion of the $P_{13}(1720)$ does not improve
the results for the $\gamma p \to p \eta$ reaction significantly.
However, this resonance provides an important contribution to reproduce the
structure exhibited by the differential cross section data in $\pi^- p \to n \eta$.
This is illustrated in Fig.~\ref{fig:pipetan5a} corresponding to the parameter
set of Table~\ref{tbl:photo5a}. In addition, the $P_{13}(1720)$ resonance also 
helps improve, to some extent, the fit quality for the total cross section at
energies above 
$W\approx 1.6$~GeV, corroborating the finding of Ref.~\citenum{GHHS03}.

\vspace{7ex}

\begin{figure*}[t!]
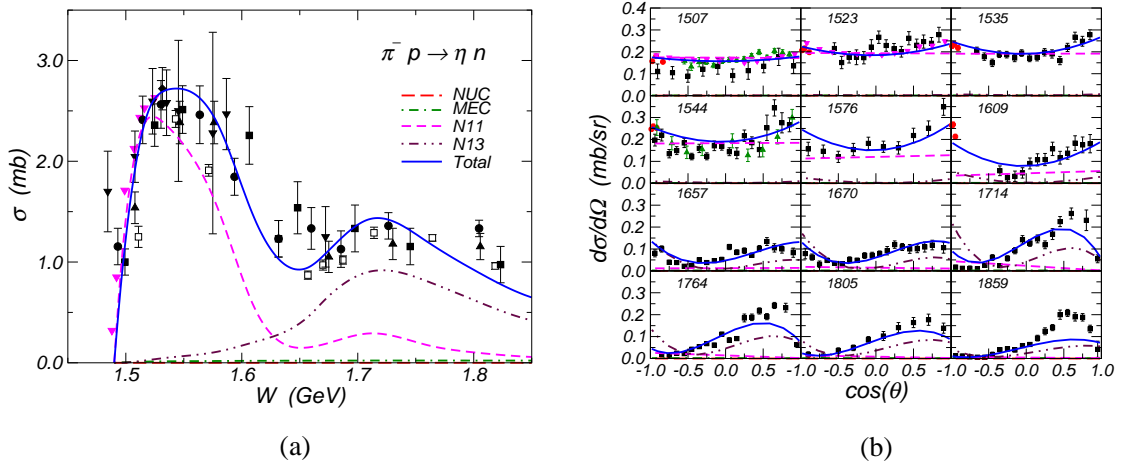
\centering
\includegraphics[width=7.0 cm]{fig12a.eps}
\hspace{0.5 cm}
\includegraphics[width=7.0 cm]{fig12b.eps}
\caption{(Color online) Same as Fig.~\ref{fig:pipetan5a} but with
another parameter set (not given here) that includes the $P_{11}(1710)$
resonance. Here, $N11$ is the sum of contributions from $S_{11}(1535)$,
$S_{11}(1650)$, and $P_{11}(1710)$.}
\label{fig:pipetanP11}
\end{figure*}

Figure~\ref{fig:pipetanP11} shows an alternative fit for the total and the
differential cross sections by using the same set of resonances as those of
Fig.~\ref{fig:pipetan5a} plus $P_{11}(1710)$.
The corresponding fit results for photoproduction are of comparable quality
to those shown in Sec.~\ref{sec:IIIA}.
As one can see from Fig.~\ref{fig:pipetanP11}(a), the small bump near
$W=1.7$ GeV in the spin-1/2 resonance contribution (dashed line) is caused
by the $P_{11}(1710)$ resonance, which makes the bump in the total contribution
(solid line) more pronounced compared to the result of
Fig.~\ref{fig:pipetan5a}(a).
The $P_{11}(1710)$ resonance seems also to affect the differential cross
section in the vicinity of $W=1670$ MeV, improving the agreement with the
data to some extent.
It is interesting to note that the chiral constituent quark-model
calculation of Ref.~\citenum{ZZHS07} shows a dominant contribution from the
$P_{11}(1710)$ resonance at an energy around $W=1700$~MeV, in contrast to the present
approach, where the dominant contribution arises from the spin-3/2 resonances.
However, the authors of Ref.~\citenum{ZZHS07} have also found that the agreement
with the measured differential cross sections in the $W=1609$--$1670$~MeV
energy range can be improved if the sign of their $P_{11}(1710)$ partial wave
amplitude is reversed and if they employ a larger total decay width of
$\Gamma_R\sim 350$ MeV.
The effect of the $P_{11}(1710)$ resonance is, then, very similar to that
found in the present calculation.
We note that the total decay width of the $P_{11}(1710)$ resonance in
Fig.~\ref{fig:pipetanP11} is $\Gamma_R=95$ MeV.

The $\pi^- p \to \eta n$ reaction has been also investigated in Ref.~\citenum{DDLSS08}
within a coupled-channel approach. There, the role of the $P$-wave resonances -- in 
particular of the $P_{11}(1440)$ and $P_{13}(1720)$ -- in the differential cross sections
has been studied. The former resonance has not been considered in the present work.

\subsection{$\bm{N N \to N N \eta}$}

\begin{figure*}[t!]
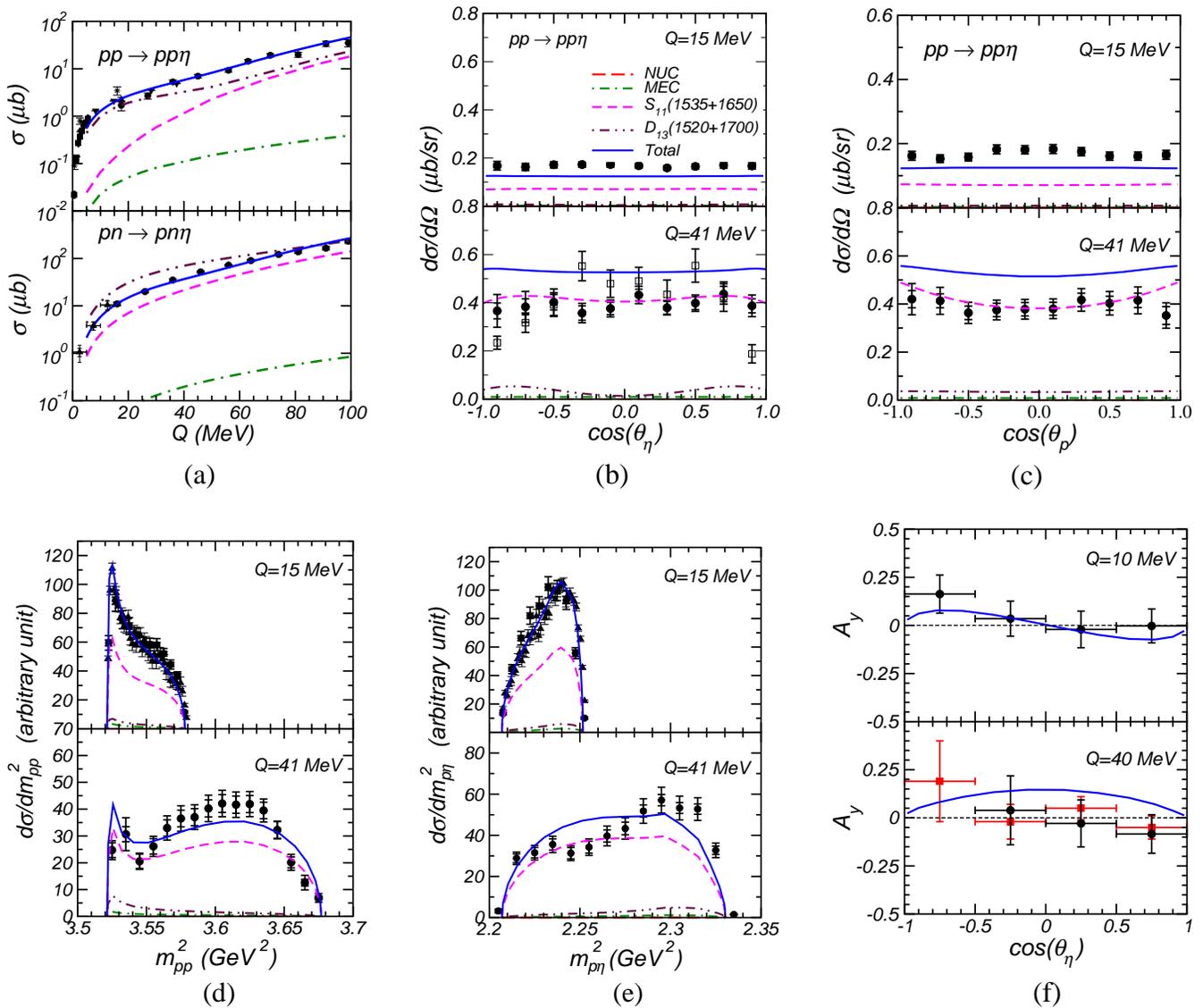
\centering
\includegraphics[width=0.3\hsize]{fig13a.eps} \qquad
\includegraphics[width=0.3\hsize]{fig13b.eps} \qquad
\includegraphics[width=0.3\hsize]{fig13c.eps} \\ \bigskip
\includegraphics[width=0.3\hsize]{fig13d.eps} \qquad
\includegraphics[width=0.3\hsize]{fig13e.eps} \qquad
\includegraphics[width=0.3\hsize]{fig13f.eps}

\caption{(Color online) Results for the reaction $N N \to N N\eta$
with the parameter set of Table~\ref{tbl:photo4c}. (a) Total cross
section as functions of the excess energy $Q$ in $pp$ and $pn$ collisions. (b)
$\eta$ angular distribution in the overall center-of-mass frame. (c) Final
proton angular distribution. (d) $pp$ invariant-mass distribution. (e) $p\eta$
invariant-mass distribution. (f) Analyzing power. In (f), only the total
contributions are shown (solid curves); the dashed curves represent the
correspond results of another parameter set (not given in this work) which
yields practically the same results for other observables considered in this
reaction. The labelings of the curves in (b, c, d, and e) are the same as in (a).
The data are from
Refs.~\citenum{NNeta-xsc-data,COSY-TOF-02,COSY11,NNeta-spin-data}.}
\label{fig:NNeta4c}
\end{figure*}

\begin{figure*}[t!]
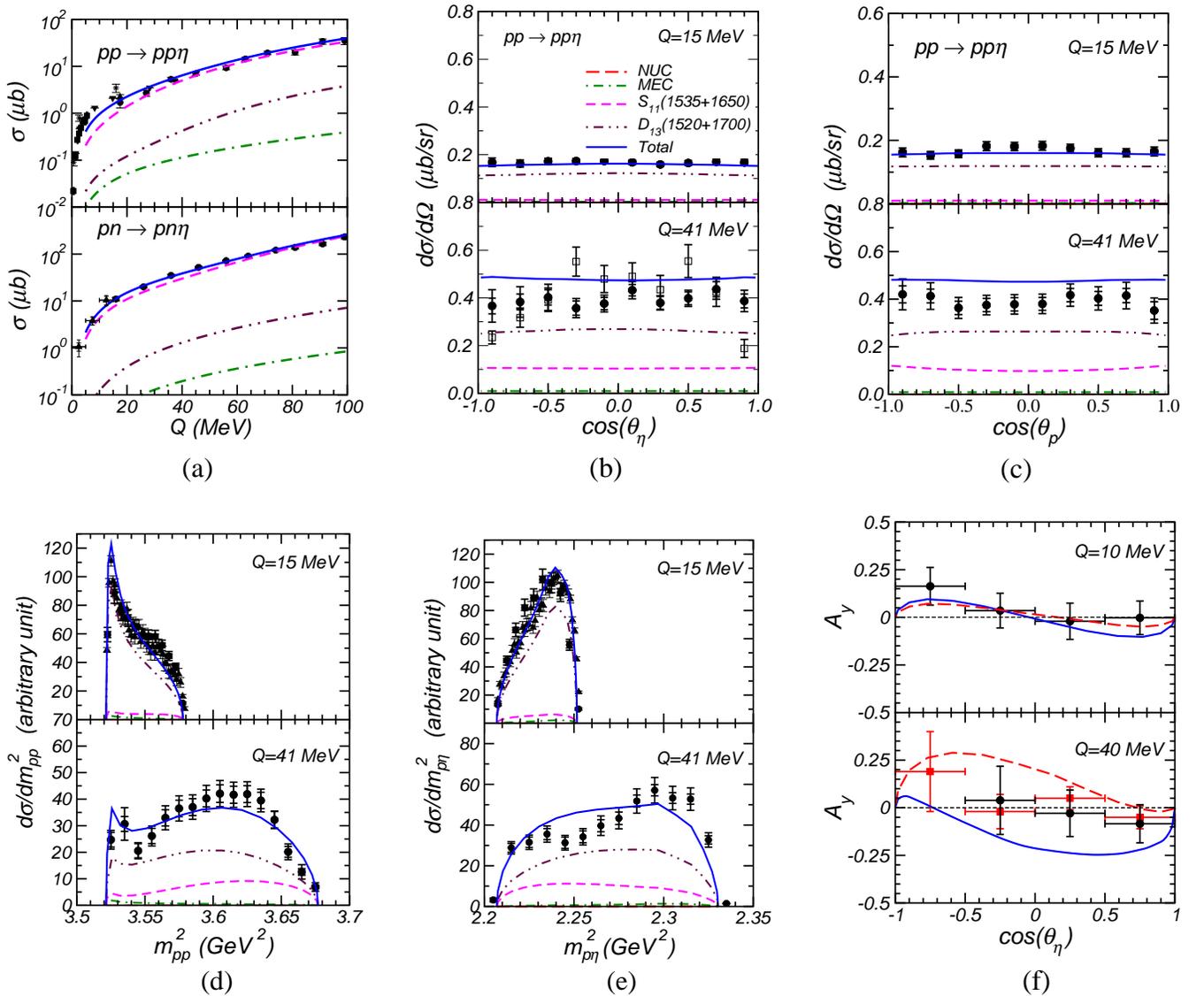
\centering
\includegraphics[width=0.3\hsize]{fig14a.eps} \qquad
\includegraphics[width=0.3\hsize]{fig14b.eps} \qquad
\includegraphics[width=0.3\hsize]{fig14c.eps} \\ \bigskip
\includegraphics[width=0.3\hsize]{fig14d.eps} \qquad
\includegraphics[width=0.3\hsize]{fig14e.eps} \qquad
\includegraphics[width=0.3\hsize]{fig14f.eps}

\caption{(Color online) Same as Fig.~\ref{fig:NNeta4c} but for the parameter set of
Table~\ref{tbl:photo4d}.} \label{fig:NNeta4d}
\end{figure*}

\begin{figure}[t!]\centering
\includegraphics[width=\columnwidth,clip=]{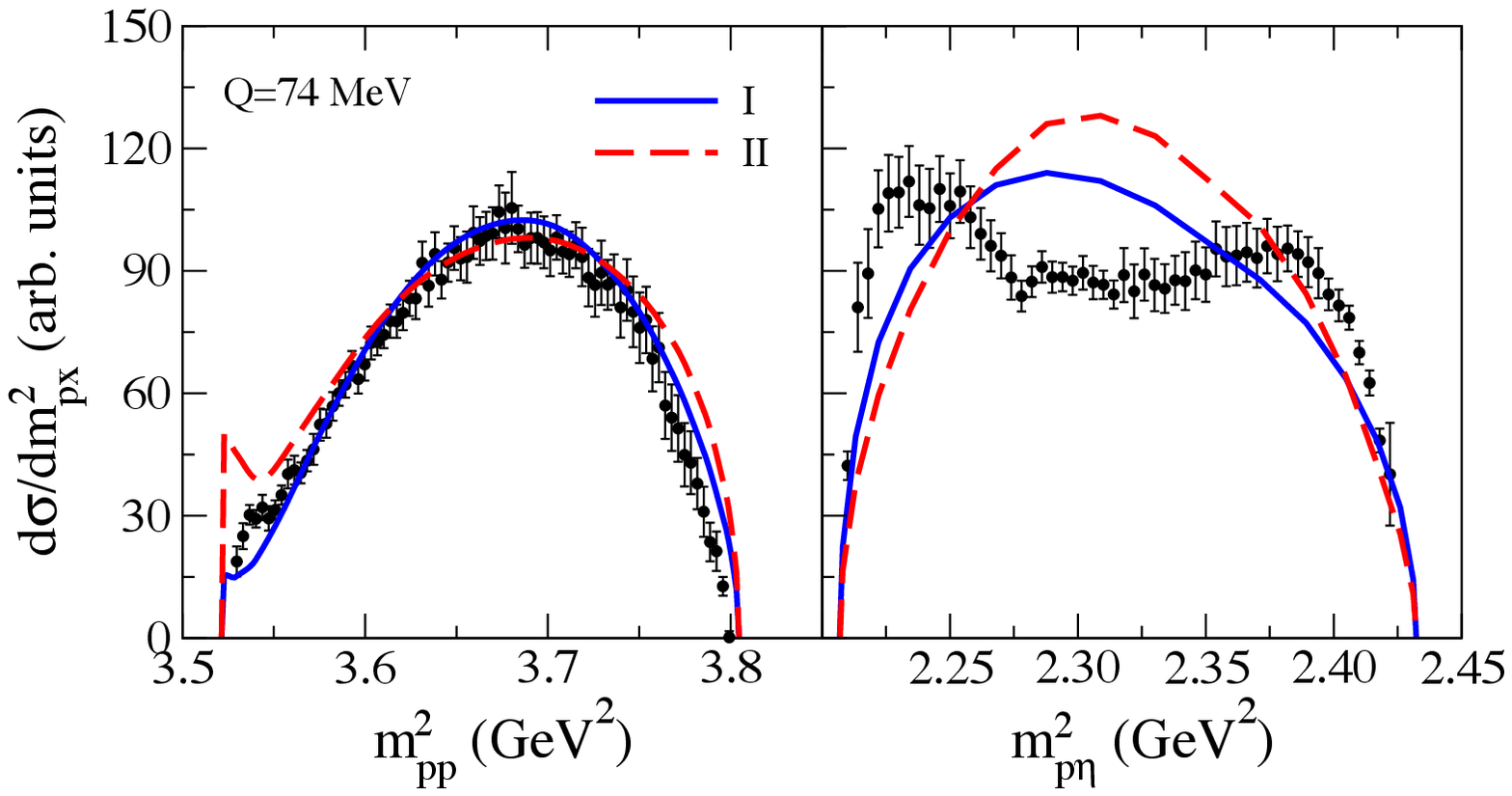}
\caption{\label{fig:NNetaQ74}(Color online)
Predictions for the invariant-mass distributions $d\sigma/dm^2_{px}$ for
$pp$ ($x=p$) and $p\eta$ ($x=\eta$) in the reaction $pp\to pp\eta$
corresponding to the excess energy of $Q=74$~MeV for the parameter sets of
Tables \ref{tbl:photo4c} (solid line) and \ref{tbl:photo4d} (dashed line).
Data are from Ref.~\citenum{CELSIUS-WASA} measured at $Q=72$ MeV. 
They have not been included in the fitting procedure.  }
\end{figure}

In this subsection, we turn our attention to the $NN \to NN\eta$ reaction.
Although the model calculation of Ref.~\citenum{NHHS03}, which is based on a strong
$P$-wave contribution, reproduces nicely the shape of the measured $pp$
invariant mass distributions, it largely underestimates the total cross
section data near the threshold.
Here, we present new results based on a combined analysis of the reactions
listed in Eq.~(\ref{eq:1}).

Shown in Fig.~\ref{fig:NNeta4c} are the results for the $NN \to NN\eta$
reaction corresponding to the parameter set of Table~\ref{tbl:photo4c}.
It can be seen that the present model reproduces rather reasonably all the
considered data for this reaction, including the energy dependence of the
total cross sections at lower energies.
This is a considerable improvement over the results of
Ref.~\citenum{NHHS03}.
However, this model still underestimates the total cross section in $pp$ collisions by
a factor of $\sim 1.5$ when the excess energy $Q$ is $Q \le 10$~MeV.
We expect the inclusion of $p\eta$ FSI and possibly the three-body
effects to resolve this discrepancy once they are properly taken into
account.
Here, the major difference of our results from those of the previous calculation of
Ref.~\citenum{NHHS03} is that we have a much stronger spin-3/2 resonance
contribution to the cross sections, especially at lower energies.
This is due to the large coupling of the $D_{13}$ resonances to the $\rho$
and the $\omega$ vector mesons, resulting from the global fitting procedure
(cf.\ Table~\ref{tbl:photo4c}).
In contrast, the $S_{11}$ resonance contribution to the cross section is
surprisingly small, especially at lower energies, which is due to the strong
destructive interference among the exchanged mesons in the excitation of
$S_{11}$ resonances.
Note that in the present calculation the resonance coupling constants
involving vector mesons, $g_{RNV}^{}$ ($V=\rho, \omega$), are basically fixed
by the $NN \to NN\eta$ reaction while the coupling constants to pseudoscalar
mesons, $g_{RNP}^{}$ ($P=\pi, \eta$), are fixed by the $\gamma p \to p \eta$
and the $\pi^- p \to n \eta$ reactions to a large extent.
We also found that the mesonic current yields a very small contribution to the
cross sections.
The nucleonic current contribution is also negligible because of the very
small $NN\eta$ coupling constant that results from the fit to the
photoproduction data, as discussed above.

It should be emphasized that it still remains to be verified whether the
dominance of the $D_{13}$ resonances discussed above is indeed true. In fact,
in spite of the present lack of information on the corresponding coupling
constants, $g^{(1,2)}_{RNV}$, the obtained values (cf.\
Table~\ref{tbl:photo4c}) may be too large to be realistic. For example, a rough
estimate of these coupling constants from the PDG helicity
amplitudes~\cite{PDG06}, in conjunction with vector-meson dominance, yields
\begin{align}
\text{for}~R=D_{13}(1520)\text{:}~~~& &
g^{(1)}_{RN\omega} &\sim -57.3,  &  g^{(2)}_{RN\omega} &\sim  82.3 ,
\nonumber \\
&&
g^{(1)}_{RN\rho}   &\sim -23.9,  &  g^{(2)}_{RN\rho}   &\sim  19.0 ,
\end{align}
and
\begin{align}
\text{for}~R=D_{13}(1700)\text{:}~~~& &
g^{(1)}_{RN\omega} &\sim -6.4,  &  g^{(2)}_{RN\omega} &\sim  9.6 ,
\nonumber \\
&&
g^{(1)}_{RN\rho}   &\sim -2.1,  &  g^{(2)}_{RN\rho}   &\sim  3.0 .
\end{align}
These values are corrected to the normalization point, at $q^2=M^2_V$, 
by writing
\begin{align}
g^{(i)}_{RNV} & \equiv  \Gamma^{(i)}_{RNV}(p'^2=M_R^2, p^2=M_N^2, q^2=M_V^2)
\nonumber \\
&=
\frac{\Gamma_{RNV}^{(i)}(p'^2=M_R^2, p^2=M_N^2, q^2=0)}
{F(p'^2=M^2_R, p^2=M^2_N, q^2=0)}
 ,
\end{align}
where $\Gamma_{RNV}^{(i)}$ is the effective hadronic coupling function, which
includes the form factor $F$ given by Eq.~(\ref{ffhadron}). We emphasize in
this context that these coupling constants cannot be determined uniquely in the
present analysis. Indeed, as shown in Fig.~\ref{fig:NNeta4d}, a scenario in
which the $S_{11}$ resonance dominates over the $D_{13}$ resonance current can
be achieved. Therefore, the consistency of these coupling constant values with
other independent reaction processes should be examined in detail. For this
purpose, vector-meson production processes, such as $NN \to NNV$, $\pi N \to N V$
and $\gamma N \to N V$, are of particular
interest~\cite{VM}.

In Ref.~\citenum{NSL02}, where the dominant $\eta$-production mechanism is the
excitation of the $S_{11}(1535)$ resonance via the pion-exchange, the analyzing
power $A_y$ exhibits a zero at $\cos(\theta_\eta) \sim 0$.
In the present calculation, where the dominant production mechanism is the
$D_{13}$ resonance excitation, the zero of $A_y$ is shifted toward backward
angles at larger $Q$.
As has been pointed out in Ref.~\citenum{NSL02}, unlike the total and the differential
cross sections, the analyzing power is very sensitive to the reaction dynamics.
In fact, the dashed curves in Fig.~\ref{fig:NNeta4c} for $A_y$ represent the
results of another parameter set (not given here) that yields practically the
same results for the other observables considered here (see also the results shown
in Fig.~\ref{fig:NNeta4d} below).
Unfortunately, the data are not accurate enough to disentangle these different
scenarios.
More accurate data will, therefore, impose more stringent constraints so as to
help distinguish different dynamics of $\eta$ production in $NN$ collisions.

The results corresponding to the parameter set of Table~\ref{tbl:photo4d} are
shown in Fig.~\ref{fig:NNeta4d}.
They are of comparable fit quality to those in Fig.~\ref{fig:NNeta4c} overall.
However, the dynamical content is quite different.
In this case, the $S_{11}$ resonance dominates over the $D_{13}$ resonance
contributions to the cross sections.
Unlike the results shown in Fig.~\ref{fig:NNeta4c}, there is no strong
destructive interference among the exchanged meson contributions to excite
the intermediate $S_{11}$ resonances.
Of course, the smallness of the $D_{13}$ resonance contributions is directly
correlated to the very small constants for the coupling of the $D_{13}$
resonances to vector mesons (cf.\ Table~\ref{tbl:photo4d}).
The analyzing power for this parameter set at $Q=40$~MeV exhibits a
qualitatively different behavior from that found in Fig.~\ref{fig:NNeta4c},
again, the quality of the current experimental data does not allow  
any definite conclusion to be drawn.

It should be emphasized that the much larger difference in the dynamics in $NN
\to NN\eta$ between the two parameter sets considered above as compared to
those in $\gamma p \to p \eta$ and $\pi^- p \to n \eta$ stems from the much
richer interference effects in the former reaction. In particular, note that,
for each meson $M$ exchanged ($\pi, \eta, \rho, \omega$), there are two
coupling vertices involved in the nucleon resonance currents (i.e., $RNM$ and
$NNM$, as shown in Fig.~\ref{fig:diagram_NNeta}). This allows for an interference among
all the exchange mesons involving the $RNM$ coupling constants, a feature that
is absent in the other two reactions. Therefore, the meson-production
reactions in $NN$ collisions, in conjunction with other basic photon- and
(two-body) hadron-induced reactions, should help in extracting the resonance
parameters to a large extent. In particular, adding the vector meson ($V$)
production channels, $NN \to NNV$, $\pi N\to N V$ and $\gamma N \to N V$,
to the list of reactions given in Eq.~(\ref{eq:1}), should impose more stringent
constraints on the resonance coupling constants.

Finally, Fig.~\ref{fig:NNetaQ74} shows predictions using the parameter sets of
Tables \ref{tbl:photo4c} and \ref{tbl:photo4d} for the invariant-mass
distributions for $pp$ and $p\eta$ at the excess energy of $Q=74$~MeV. 
The results are compared with the recent CELSIUS-WASA data~\cite{CELSIUS-WASA}
measured at $Q=72$~MeV. Note that these data have not been included in the 
present fit. 
We see that, while the $pp$ invariant mass distribution is well reproduced 
except in the small $m_{pp}$ region, the $p\eta$ invariant mass distribution 
shows big differences between the prediction of the present model and
the measured data, indicating a deficiency in the model. Note that the present model
does not account for the $\eta N$ FSI. Further investigation is required to 
identify the origin of the discrepancies in both the $pp$ and the $p\eta$ invariant
mass distributions.

\section{Summary}
\label{sec:summary}

In the present work, a combined analysis of the reactions $\gamma p \to p
\eta$, $\pi^- p \to n \eta$, $pp \to pp\eta$, and $pn \to pn\eta$ has been
carried out within a relativistic meson-exchange model of hadronic
interactions.
Both the $\gamma p \to p \eta$ and the $\pi^- p \to n \eta$ reactions have been
treated in the tree-level approximation with the former reaction containing
a generalized contact current that ensures gauge invariance of the reaction
amplitude.
The $NN \to NN\eta$ reaction has been treated in the DWBA approximation with
the explicit treatment of the $NN$ ISI and FSI.
The free parameters of the model, especially the nucleon resonance parameters,
are then fixed in a global fitting procedure.

Overall, the photoproduction data can be described reasonably well with the
inclusion of the well-established $S_{11}(1535)$, $S_{11}(1650)$,
$D_{13}(1520)$, and $D_{13}(1700)$ resonances as the minimally required set of
resonances to achieve a reasonable fit to the currently available data.
The inclusion of additional well-known resonances in the same mass region
[including the spin-5/2 $D_{15}(1675)$ and $F_{15}(1680)$ resonances] does
not further improve the quality of the  overall description of the data.
The measured angular distributions at higher energies and backward angles are
compatible with a vanishing $NN\eta$ coupling constant.
However, in order to extract this coupling constant unambiguously within an
approach of the type pursued here, one needs to go beyond the resonance region
to avoid possible interference effects.
On the other hand, as we have seen in the results of
Fig.~\ref{fig:photo52b_he_sb}, the beam asymmetry can impose constraints on
the $NN\eta$ coupling constant that are much more stringent than the
differential cross sections because of the interference between the
nucleonic and the resonance currents.
One difficulty of the present calculation is to reproduce the
$\sin(2\theta_\eta)$ dependence exhibited by the measured target asymmetry
near threshold, and this shows that further investigations are necessary to
understand better the production mechanism of this reaction.

Our model can also explain the available data on $\pi^- p \to n \eta$
reasonably well, at least for energies not too far from the threshold.
At higher energies the total cross section is underestimated due to the lack
of the $N\pi\pi$ contribution via the coupled channel.

Our model also describes the $NN \to NN\eta$ reaction data rather well. 
The problem of the underestimation of the total cross section
near threshold~\cite{NHHS03} has been cured to a large extent in the present 
approach. However, we emphasize that the scenario of strong coupling of the $D_{13}$ 
resonance to the $NV$ channels ($V=\rho,\omega$) still remains to be verified, for 
the $RNV$ couplings cannot be fixed unambiguously in the present study. In this 
connection, vector meson production reactions, such as the $\gamma N \to NV$,
$\pi N \to NV$, $N N \to N N V$, should be investigated.
We also verified that the analyzing power is sensitive to the
reaction dynamics. Unfortunately, however, the currently available data are not
accurate enough to unambiguously distinguish different dynamical contributions.

As we have illustrated with some selected examples, the present approach is
unable to determine a unique set of (resonance) parameter values.
In fact, it was shown that different parameter sets that describe the data
equally well lead to results that exhibit quite different reaction dynamics.
This is especially true for the $NN \to NN\eta$ reaction,
where we found quite different values of the $RNV$ coupling constants.
As mentioned above, the inclusion of vector-meson production reactions 
into the combined analysis
should help constrain those coupling constants.
Consequently, our study reveals that one must be cautious in interpreting the
resonance parameters extracted from these kinds of analyses, especially, if only
a single reaction process is considered.
It is clear that in order to extract more accurate information on the nucleon
resonances, one must combine the investigation of hadron-induced meson
production with the corresponding photon-induced reactions.
To help in this, ample data sets are now available for photo- and
electro-production processes from various accelerator facilities, including the
Thomas Jefferson National Accelerator Facility, SPring-8, CB/ELSA, GRAAL, etc.
By contrast, data for hadron-induced production processes are much more limited,
and we clearly need more of them.

In this connection, we note that the progress in the study of meson-production
processes in $NN$ collisions, both experimentally and theoretically, has
reached such a level that it allows us to address certain concrete physics
issues, especially, when they are investigated in conjunction with other
independent reactions.
This has been illustrated in the present work for the specific case of $\eta$
production, where some information on nucleon resonances can be extracted.
In particular, the consideration of meson-production processes in $NN$
collisions, in conjunction with photon- and (two-body) hadron-induced reactions
aimed at a resonance parameters extraction, should help impose more stringent
constraints on these parameters.
As pointed out in the last subsection, meson production in $NN$ collisions
exhibits much richer interference effects, a feature that is absent in more
basic two-body reactions.
Furthermore, the inclusion of this reaction in the resonance-parameter
extraction is especially relevant because the existing data for meson
production (other than for the pion) in two-body hadronic reactions are rather
scarce and of relatively low accuracy.
Currently, there exist only very limited efforts to improve or extend the
corresponding database~\cite{CB05}.
On the other hand, the available data on meson production in $NN$ collisions
are much more accurate; moreover, the corresponding database can be and
is being expanded, especially at the COSY facility.

\begin{acknowledgments}

This work was supported by the Forschungs\-zentrum J{\"u}lich under
FFE Grant No.\ 41445282 and by the Basic Science Research Program through 
the National Research Foundation of Korea (NRF) funded by the 
Ministry of Education, Science and Technology (Grant \mbox{No.} 2010-0009381).

\end{acknowledgments}

\appendix*

\section{}

In this Appendix, we give the ingredients that define our models
described in Sec.~\ref{sec:model}.
Throughout this paper, we use the notation $N$ and $R$ for the nucleon
and the nucleon resonance fields, respectively;
$M_B$ denotes the mass of the baryon $B$ ($=N,R$).
We also use $S$ $(=\sigma, \vec a_0)$, $P$ $(=\eta, \vec\pi)$, and
$V_\mu$ $(=\omega_\mu, \vec\rho_\mu)$ to denote the scalar, pseudoscalar,
and vector meson fields, respectively.
The vector notation refers to the isospin space.
For isovector mesons, $S \equiv \vec S \cdot \vec \tau$,
$P \equiv \vec P \cdot \vec \tau$, and
$V_\mu \equiv \vec V_\mu \cdot \vec \tau$.
The mass of the meson $M$ is denoted by $M_M$ $(M=S,P,V)$.
The photon field is denoted by $A_\mu$.
We define $V^{\mu\nu}\equiv\partial^\mu V^\nu-\partial^\nu V^\mu$ and
$F^{\mu\nu}\equiv\partial^\mu A^\nu-\partial^\nu A^\mu$.

We use the superscript $j^P$ in the Lagrangian densities (${\cal L}^{(j^P)}$)
involving the nucleon resonance $R$ to denote the spin-parity $j^P$ of
that resonance.
Furthermore, for convenience, we define
\begin{equation}
\Gamma^{(+)} \equiv \gamma_5
\qquad\text{and}\qquad
\Gamma^{(-)} \equiv 1 ~.
\end{equation}

\subsection{Hadronic Interaction Lagrangians}

The following interaction Lagrangian densities describe the hadronic vertices.
The Lagrangians for meson-nucleon interactions are
%
\begin{align}
{\cal L}_{NNS} &=  g_{NNS}^{} \,\bar N N S ,
\label{NNS}
\displaybreak[0]\\
{\cal L}_{NNP} &=
- g_{NNP}^{}\, \bar N \left\{
\Gamma^{(+)} \left[ i\lambda +
\frac{1 - \lambda}{2M_N}\, \fs{\partial} \right] P \right\}
N ,
\label{NNP} \displaybreak[0]\\
{\cal L}_{NNV} &=  - g_{NNV}^{} \, \bar N \left\{
\left[ \gamma^\mu - \kappa_V^{} \frac{\sigma^{\mu\nu}\partial_\nu}{2M_N}
\right] V_\mu \right\} N ,
\label{NNV}
\end{align}
where the parameter $\lambda$ was introduced in $\mathcal{L}_{NNP}$
to interpolate between
the pseudovector ($\lambda=0$) and the pseudoscalar ($\lambda=1$)
couplings.
We use the meson-nucleon-nucleon coupling constants in the above 
Lagrangians as in Ref.~\citenum{NSL02}: namely, $g_{NN\omega}^{} = 17.47$,
$\kappa_\omega^{} = 0$, $g_{NN\rho}^{} = 3.36$, $\kappa_\rho = 6.1$, and
$g_{NNa_0^{}}^{} = 5.59$.

\begin{widetext}
%
The effective Lagrangians describing the interactions of the nucleon
resonance with the nucleon and the pseudoscalar meson $P$ or vector meson
$V$ read
\begin{align}
{\cal L}^{(\frac{1}{2}^\pm)}_{RNP}
&=
\mp g_{RNP}^{}
\bar R  \left\{ \Gamma^{(\pm)} \left[ i\lambda +
\frac{1 - \lambda}{M_R\pm M_N}\, \fs{\partial}
\right]P\right\} N
+  \hc ,
\label{PNR}
\displaybreak[0]\\
{\cal L}^{(\frac{1}{2}^\pm)}_{RNV}
&=  - \frac{1}{2M_N} \bar R \Gamma^{(\mp)}
\left\{ \left[g_{RNV}^{}\left(
\frac{\gamma_\mu\partial^2}{M_R\mp M_N} - i\partial_\mu
\right)  - f_{RNV}^{}\sigma_{\mu\nu}\partial^\nu \right] V^\mu \right\} N
+ \hc ,
\label{VNR}
\end{align}
for a resonance of spin-$\frac12$. The Lagrangian in Eq.~(\ref{PNR}) contains the
pseudoscalar-pseudovector mixing parameter $\lambda$, similar to
Eq.~(\ref{NNP}).
%
%
For spin-$\frac32$ resonances, we use
\begin{align}
\mathcal{L}^{(\frac{3}{2}^\pm)}_{RNP} &= \frac{g_{RNP}^{}}{M_P}
{\bar{R}^\mu} \Theta_{\mu\nu}(z) \Gamma^{(\pm)} (\partial^\nu P) N +\hc ,
 \label{PNR32}
\displaybreak[0]\\
\mathcal{L}^{(\frac{3}{2}^\pm)}_{RNV} &= -i\frac{g^{(1)}_{RNV}}{2M_N}
{\bar{R}^\beta} \Theta_{\beta\mu} \Gamma^{(\pm)}\gamma_\nu V^{\mu\nu} N
- \frac{g^{(2)}_{RNV}}{4M^2_N}
\bar{R}^\beta \Theta_{\beta\mu} \Gamma^{(\pm)} V^{\mu\nu} \partial_\nu N
\mp \frac{g^{(3)}_{RNV}}{4M^2_N} \bar{R}^\beta \Theta_{\beta\mu}
\Gamma^{(\pm)} \left(\partial_\nu V^{\mu\nu}\right) N
+ \hc ,
\label{VNR32}
\end{align}
where the coupling tensor is
$\Theta_{\mu\nu}=g_{\mu\nu}-(z+\frac{1}{2})\gamma_\mu\gamma_\nu$
and we take the off-shell parameter $z=-\frac{1}{2}$ for simplicity.
%
For the interaction Lagrangians of spin-$\frac52$ resonances, we
use
\begin{align}
\mathcal{L}^{(\frac{5}{2}^\pm)}_{RNP} &= i\frac{g_{RNP}^{}}{M_P^2}
\bar{N} \Gamma^{(\pm)} \left(\partial^\mu\partial^\nu P \right) R_{\mu\nu}
+ \hc ,
 \label{PNR52}
\displaybreak[0]\\
\mathcal{L}^{(\frac{5}{2}^\pm)}_{RNV}&=
\frac{g^{(1)}_{RNV}}{(2M_N)^2}
\bar{N} \gamma_\nu\Gamma^{(\mp)}\left(\partial^\alpha V^{\mu\nu}\right)
R_{\mu\alpha}
-i \frac{g^{(2)}_{RNV}}{(2M_N)^3}
(\partial_\nu \bar{N}) \Gamma^{(\mp)}
\left(\partial^\alpha V^{\mu\nu}\right) R_{\mu\alpha}
+ i \frac{g^{(3)}_{RNV}}{(2M_N)^3}
\bar{N} \Gamma^{(\mp)}
\left(\partial^\alpha\partial_\nu V^{\mu\nu}\right) R_{\mu\alpha}
\nonumber\\ &\quad \mbox{}
+\hc ,
\label{VNR52}
\end{align}
\end{widetext}

The hadronic interaction Lagrangians among mesons are
%
\begin{subequations}
\begin{align}
{\cal L}_{\rho\rho\eta} &=  - \frac{g_{\rho\rho\eta}^{}}{2M_\rho}
\varepsilon_{\alpha\beta\nu\mu} (\partial^\alpha \vec{\rho}^{\,\beta}) \cdot
(\partial^\nu \vec{\rho}^{\,\mu}) \eta ~,
\displaybreak[0]\\
{\cal L}_{\omega\omega\eta} &=  - \frac{g_{\omega\omega\eta}^{}}{2M_\omega}
\varepsilon_{\alpha\beta\nu\mu} (\partial^\alpha \omega^\beta)
(\partial^\nu \omega^\mu) \eta ~,
\displaybreak[0]\\
{\cal L}_{\pi\eta a_0^{}} &=
\frac{g_{\pi\eta a_0^{}}^{}}{\sqrt{M_\pi M_\eta}}
(\partial_\mu\eta)(\partial^\mu\vec\pi)\cdot\vec a_0^{}~,
\label{PPS}
\end{align}
\end{subequations}
with the convention $\varepsilon_{0123}^{} = -1$ for the Levi-Civita
antisymmetric tensor.
We follow Refs.~\citenum{NSL02} and \citenum{NH04-NH05} for the coupling constant values at the
$NNM$ vertices above, except for the coupling constant $g_{NN\eta}^{}$,
which is treated as a free parameter in the present work.
The coupling constants of the $RNM$ interaction Lagrangians, as well as
the resonance masses $M_R$, are free parameters to be adjusted to
reproduce the data.
For the other coupling constants, following Ref.~\citenum{NSL02}, we use
$g_{\eta\rho\rho}^{}=4.94$, $g_{\eta\omega\omega}^{}=4.84$,
and $g_{\eta\pi a_0}^{}=1.81$.

\subsection{Electromagnetic Interaction Lagrangians}

The electromagnetic vertices are calculated from the Lagrangian
densities given below.
The electromagnetic interaction of the nucleon reads
\begin{equation}
{\cal L}_{NN\gamma} =  -e \bar N \left\{
\left[\hat e \gamma^\mu - \hat\kappa\frac{\sigma^{\mu\nu}\partial_\nu}{2M_N}
\right] A_\mu \right\} N ,
\label{NNgamma}
\end{equation}
where $e$ stands for the elementary charge unit, and
$\hat e \equiv (1 + \tau_3)/2$ and
$\hat\kappa \equiv \kappa_p^{}(1 + \tau_3)/2 + \kappa_n^{}(1 - \tau_3)/2$,
with the anomalous magnetic moments $\kappa_p^{}=1.739$ for the proton
and $\kappa_n^{}=-1.931$ for the neutron.
\begin{widetext}
The photo-transition Lagrangians of resonances into the nucleon are
%
\begin{align}
\mathcal{L}^{(\frac{1}{2}^\pm)}_{RN\gamma}
& =  e \frac{g_{RN\gamma}^{(1)}}{2M_N} \bar R \Gamma^{(\mp)}
\sigma_{\mu\nu}\partial^\nu A^\mu N +\hc ,
\label{RNgamma}
\displaybreak[0]\\
%
\mathcal{L}^{(\frac{3}{2}^\pm)}_{RN\gamma}&=
-ie\frac{g^{(1)}_{RN\gamma}}{2M_N}
{\bar{R}^\beta} \Theta_{\beta\mu} \Gamma^{(\pm)}\gamma_\nu F^{\mu\nu} N
- e\frac{g^{(2)}_{RN\gamma}}{4M^2_N}
\bar{R}^\beta \Theta_{\beta\mu} \Gamma^{(\pm)} F^{\mu\nu} \partial_\nu N
+ \hc ,
\label{gammaNR32}
\displaybreak[0]\\
%
\mathcal{L}^{(\frac{5}{2}^\pm)}_{RN\gamma}&=
e\frac{g^{(1)}_{RN\gamma}}{(2M_N)^2}
\bar{N} \gamma_\nu\Gamma^{(\mp)}\left(\partial^\alpha F^{\mu\nu}\right)
R_{\mu\alpha}
-ie \frac{g^{(2)}_{RN\gamma}}{(2M_N)^3}
(\partial_\nu \bar{N}) \Gamma^{(\mp)}
\left(\partial^\alpha F^{\mu\nu}\right) R_{\mu\alpha}
+ \hc ,
\label{NN*52gamma}
\end{align}
\end{widetext}
and the photo-transition Lagrangian between mesons are
%
\begin{subequations}
\begin{align}
{\cal L}_{\eta\rho\gamma} &=  - e\frac{g_{\eta\rho\gamma}^{}}{M_\rho}
\varepsilon_{\alpha\beta\nu\mu} (\partial^\alpha \rho_0^{\,\beta})
(\partial^\nu A^{\,\mu}) \eta ,
\displaybreak[0]\\
{\cal L}_{\eta\omega\gamma} &= - e\frac{g_{\eta\omega\gamma}^{}}{M_\omega}
\varepsilon_{\alpha\beta\nu\mu} (\partial^\alpha \omega^\beta)
(\partial^\nu A^\mu) \eta .
\end{align}
\end{subequations}

The electromagnetic coupling constants involving nucleon resonances in
the above Lagrangians are free parameters to be adjusted to fit the
pertinent experimental data.
The coupling constants $g_{\eta v \gamma}^{}$  $(v=\rho,\omega)$ in the
above equations are determined from a systematic analysis of the
pseudoscalar and the vector meson radiative decays~\cite{Durso87,NSL02}.
That analysis leads to $g_{\eta\rho\gamma}^{}=1.44$
and $g_{\eta\omega\gamma}^{}=0.47$.
Their signs are inferred also from $SU(3)$ flavor symmetry considerations
in conjunction with the sign of the coupling constant
$g_{\pi V \gamma}^{}$ determined from a study of pion photoproduction
in the 1~GeV energy region~\cite{GM93}.

\subsection{Form Factors}

Each vertex obtained from the interaction Lagrangians given
above is multiplied by a phenomenological cutoff function
\begin{equation}
F(p'^2, p^2, q^2) = F_B(p'^2) F_B(p^2) F_M(q^2)  ,
\label{ffhadron}
\end{equation}
where $p'$ and $p$ denote the four-momenta of the two baryons, and $q$ is the
four-momentum of the meson at the three-point vertex.
Here, we use
\begin{equation}
F_B(x) =\frac{\Lambda_B^4}{\Lambda_B^4+\left( x-M_B^2 \right)^2}  ,
\label{ffbaryon}
\end{equation}
where the cutoff $\Lambda_B=1200$~MeV is taken to be the same for all the baryons,
and $F_M(q^2)$ is given by
\begin{equation}
F_M(q^2) = \left( \frac{\Lambda_M^2-M_M^2}{\Lambda_M^2-q^2} \right)^n  ,
\label{ffmeson}
\end{equation}
with $n=1$ for a scalar or a pseudoscalar meson and $n=2$ for a vector meson.
The values of $\Lambda_M$ are taken to be the same as those used in
Ref.~\citenum{NSL02}.

The electromagnetic $\eta v\gamma$ vertex is multiplied by the form
factor $G_v(q^2)$, which describes the off-shell behavior of the
intermediate vector meson with squared momentum transfer $q^2$
(cf.\ the fourth diagram in Fig.~\ref{fig:diagram_photo}).
In general, we use the dipole form
\begin{equation}
G_v(q^2) = \left( \frac{\Lambda^{*2}_v}{\Lambda^{*2}_v-q^2} \right)^2  ,
\label{ffgamma}
\end{equation}
where the cutoff $\Lambda^*_v$, taken to be the same for both
$\rho$ and $\omega$, is a free parameter.

\subsection{Propagators}

The calculation of the Feynman diagrams displayed in
Figs.~\ref{fig:diagram_photo}, \ref{fig:diagram_pieta}, and
\ref{fig:diagram_NNeta} requires the corresponding baryon and meson
propagators, whose explicit forms are given here:
\begin{align}
S(p) &= \frac{1}{\fs{p} - M_N +i\epsilon}~  ,
\displaybreak[0]\\
\Delta(q^2) &= \frac{1}{q^2 - M_M^2 + i\epsilon}~  , \displaybreak[0] \\
D^{\mu\nu}(q) &= \frac{-g^{\mu\nu}
+ {q^\mu q^\nu}/{M_V^2}}{q^2 - M_V^2 + i\epsilon}~   ,
\label{NM-prop}
\end{align}
where $S(p)$ is the nucleon propagator with the nucleon four-momentum $p$,
$\Delta(q^2)$ is the scalar or pseudoscalar meson propagator with
four-momentum $q$, and $D^{\mu\nu}(q)$ is the vector meson propagator.
For a spin-1/2 resonance propagator, we use the ansatz
\begin{align}
S_{1/2}(p) &= \frac{1}{\fs{p}-M_R+\frac{i}{2}\Gamma}
\nonumber \\
&= \frac{\fs{p}+M_R}{p^2-M_R^2+\frac{i}{2}(\fs{p}+M_R)\Gamma} ~,
\label{eq:N12-prop}
\end{align}
where $\Gamma$ is the energy dependent resonance width.
For spin-3/2, the resonant propagator reads in a schematic matrix notation,
\begin{equation}
S_{3/2}(p)=\left[(\fs{p}-M_R)g-i\frac{\Delta}{2}\Gamma\right]^{-1}\Delta ~,
\label{N32-prop}
\end{equation}
where all indices are suppressed; i.e., $g$ is the metric tensor and
$\Delta$ is the Rarita-Schwinger tensor written in full detail as
\begin{equation}
\Delta^{\mu\nu}_{\beta\alpha}=
-g^{\mu\nu}\delta_{\beta\alpha}+\frac{1}{3}\gamma^\mu_{\beta\varepsilon}
\gamma^\nu_{\varepsilon\alpha}
  + \frac{2p^\mu p^\nu}{3M_R^2}\delta_{\beta\alpha}
+\frac{\gamma^\mu_{\beta\alpha}
p^\nu-p^\mu\gamma^\nu_{\beta\alpha}}{3M_R} ,
\label{eq:RStensor}
\end{equation}
where $\alpha$, $\beta$, and $\varepsilon$ enumerate the four indices
of the $\gamma$-matrix components
(summation over $\varepsilon$ is implied).
The inversion in Eq.~(\ref{N32-prop}) is to be understood on the full
16-dimensional space of the four Lorentz indices and the four components
of the gamma matrices.
The motivation for the ansatz in Eq.~(\ref{N32-prop}) and the technical
details on how to perform this inversion are given in
Ref.~\citenum{NH04-NH05}.

Similarly, the propagator for a spin-5/2 resonance is given by
\begin{equation}
S_{5/2}(p)=\left[(\fs{p}-M_R)g-i\frac{G}{2}\Gamma\right]^{-1}G .
\label{N52-prop}
\end{equation}
where
\begin{align}
G &\equiv G^{\mu\nu ; \rho\tau}_{\beta\alpha} = \left[
\frac{1}{2}\left(\bar{g}^{\mu\rho}\bar{g}^{\nu\tau} +
\bar{g}^{\mu\tau}\bar{g}^{\nu\rho}\right)
 - \frac{1}{5}\bar{g}^{\mu\nu}\bar{g}^{\rho\tau} \right]
\delta_{\beta\alpha} \nonumber \\
& - \frac{1}{10}\left(
\bar{g}^{\mu\rho}\bar{\gamma}^\nu_{\beta\varepsilon}
\bar{\gamma}^\tau_{\varepsilon\alpha} +
\bar{g}^{\mu\tau}\bar{\gamma}^\nu_{\beta\varepsilon}
\bar{\gamma}^\rho_{\varepsilon\alpha} +
\bar{g}^{\nu\rho}\bar{\gamma}^\mu_{\beta\varepsilon}
\bar{\gamma}^\tau_{\varepsilon\alpha} +
\bar{g}^{\nu\tau}\bar{\gamma}^\mu_{\beta\varepsilon}
\bar{\gamma}^\rho_{\varepsilon\alpha} \right) ,
\label{eq:52tensor}
\end{align}
with
\begin{equation}
\bar{g}^{\mu\nu} \equiv g^{\mu\nu} - \frac{p^\mu p^\nu}{M_R^2},
\qquad
\bar{\gamma}^\mu \equiv \gamma^\mu - \frac{p^\mu \fs{p}}{M_R^2}  .
\end{equation}
The above spin-5/2 propagator is a variant of the one given in
Ref.~\citenum{52prop}.

We write the resonance width $\Gamma$ as a function of $W=\sqrt{s}$
according to
\begin{equation}
\Gamma(W) = \Gamma_R\left[\sum_{i=1}^N \beta_i^{} \hat{\Gamma}_i(W) +
\sum_{j=1}^{N_\gamma} \gamma_j^{} \hat{\Gamma}_{\gamma_j}(W)\right] ,
\end{equation}
where the sums over $i$ and $j$, respectively, account for decays of the
resonance into $N$ and two- or three-hadron channels and into
$N_\gamma$ radiative decay channels.
The total static resonance width is denoted by $\Gamma_R$, and
the numerical factors $\beta_i^{}$ and $\gamma_j^{}$
(with $0\le \beta_i^{},\gamma_j^{} \le 1$) describe the branching ratios
into the corresponding decay channel; i.e.,
\begin{equation}
\sum_{i=1}^N \beta_i^{} +\sum_{j=1}^{N_\gamma}\gamma_j^{} =1 .
\label{eq:brsum}
\end{equation}
We parametrize the width functions $\hat{\Gamma}_i$ and
$\hat{\Gamma}_{\gamma_j^{}}$ (which are both normalized to $1$ at $W=M_R$) as
given in Ref.~\citenum{NH04-NH05} to provide the correct respective threshold behaviors.

\end{document}